\pdfoutput=1 
\documentclass[cernpreprint,english]{na61doc}
\usepackage[T1]{fontenc}
\usepackage[labelfont=bf]{caption}
\usepackage{mathptmx} 
\usepackage{enumerate}
\usepackage{mathtools} 
\usepackage{amssymb}
\usepackage{placeins}
\usepackage{appendix}
\usepackage[utf8]{inputenc}
\usepackage{chngcntr}
\usepackage{microtype}
\usepackage{lineno}
\usepackage{color}
\usepackage{epstopdf}
\usepackage{colortbl}
\usepackage{tabularx}
\usepackage{multirow}

\usepackage{subcaption}
\usepackage{booktabs}
\usepackage{url}
\usepackage{cite}
\usepackage{caption}

\usepackage{hyperref}

\usepackage{svg}
\usepackage{tikz}
\usetikzlibrary{patterns}
\usetikzlibrary{plotmarks}
\usepackage{pgfplots}
\usepackage{longtable}

\usepackage{comment}
\usepackage{amsmath}
\usepackage{amsbsy}

\usepackage{array}
\newcolumntype{g}{>{\global\let\currentrowstyle\relax}}
\newcolumntype{^}{>{\currentrowstyle}}

\definecolor{tab_blue}{HTML}{1f77b4}
\definecolor{tab_orange}{HTML}{ff7f0e}
\definecolor{tab_green}{HTML}{2ca02c}
\definecolor{tab_red}{HTML}{d62728}
\definecolor{tab_violet}{HTML}{9467bd}
\definecolor{tab_brown}{HTML}{8c564b}
\definecolor{tab_pink}{HTML}{e377c2}
\definecolor{tab_gray}{HTML}{7f7f7f}
\definecolor{tab_yellow}{HTML}{bcbd22}
\definecolor{tab_cyan}{HTML}{17becf}

\definecolor{darkred}{rgb}{0.5,0,0}
\definecolor{darkblue}{rgb}{0,0,0.5}
\definecolor{firebrick}{rgb}{0.75,0.125,0.125}
\definecolor{darkgreen}{rgb}{0,0.5,0}
\definecolor{kPink+2}{RGB}{204,102,153}
\definecolor{kOrange+8}{RGB}{255,102,51}
\definecolor{kGreen+2}{RGB}{0,153,0}
\definecolor{kCyan+2}{RGB}{0,153,153}
\definecolor{kBlue+2}{RGB}{0,0,153}
\definecolor{kRed+1}{RGB}{204,0,0}
\definecolor{kBlue}{RGB}{0,0,204}
\definecolor{kBlue-9}{RGB}{153,153,255}
\definecolor{kGreen}{RGB}{0,153,0}
\definecolor{kRed}{RGB}{204,0,0}
\definecolor{kCyan}{RGB}{51,204,204}
\definecolor{kMagenta}{RGB}{153,0,153}
\definecolor{kPink}{RGB}{204,0,102}
\definecolor{kGray}{RGB}{204,204,204}
\definecolor{kBlack}{RGB}{0,0,0}

\definecolor{kRed+3}{RGB}{102,0,0}
\definecolor{kRed+2}{RGB}{153,0,0}
\definecolor{kRed-4}{RGB}{255,51,51}
\definecolor{kRed-7}{RGB}{255,102,102}
\definecolor{kRed-9}{RGB}{255,153,153}
\definecolor{redShading}{RGB}{229,127,127}
\definecolor{kBlue+1}{RGB}{0,0,204}

\newcommand{\eV}{\ensuremath{\mbox{e\kern-0.1em V}}\xspace}
\newcommand{\GeV}{\ensuremath{\mbox{Ge\kern-0.1em V}}\xspace}
\newcommand{\MeV}{\ensuremath{\mbox{Me\kern-0.1em V}}\xspace}
\newcommand{\GeVc}{\ensuremath{\mbox{Ge\kern-0.1em V}\!/\!c}\xspace}
\newcommand{\GeVcc}{\ensuremath{\mbox{Ge\kern-0.1em V}\!/\!c^2}\xspace}
\newcommand{\AGeV}{\ensuremath{A\,\mbox{Ge\kern-0.1em V}}\xspace}
\newcommand{\AGeVc}{\ensuremath{A\,\mbox{Ge\kern-0.1em V}\!/\!c}\xspace}
\newcommand{\MeVc}{\ensuremath{\mbox{Me\kern-0.1em V}/c}\xspace}

\newcommand{\mm}{\ensuremath{\mbox{mm}}\xspace}

\newcommand{\dd}{\ensuremath{{\mathrm{d}}}\xspace}
\newcommand{\dedx}{\ensuremath{\dd E\!/\!\dd x}\xspace}

\newcommand{\pt}{\ensuremath{p_{T}}\xspace}
\newcommand{\y}{\ensuremath{{y}}\xspace}

\newcommand{\Ks}{\ensuremath{K^0_S}\xspace}

\newcommand{\pim}{\ensuremath{\pi^-}\xspace}
\newcommand{\pip}{\ensuremath{\pi^+}\xspace}

\newcommand{\snn}{\sqrt{s_{NN}}}

\newcommand{\CernVM}{\textsc{Cern\-\kern-0.05emVM}\xspace}
\newcommand{\NASixtyOne}{NA61\slash SHINE\xspace}%

\ShineTitle{
Evidence of isospin-symmetry violation in high-energy collisions of atomic nuclei 
}

\PreprintIdNumber{CERN-EP-2023-283}

\ShineAbstract{
Strong interactions preserve an approximate isospin symmetry between up ($u$) and down ($d$) quarks, part of the more general flavor symmetry. In the case of $K$ meson production, if this isospin symmetry were exact, it would result in equal numbers of charged ($K^+$ and $K^-$) and neutral ($K^0$ and $\overline K^{\,0}$) mesons produced in collisions of isospin-symmetric atomic nuclei. Here, we report results on the relative abundance of charged over neutral $K$ meson production in argon and scandium nuclei collisions at a center-of-mass energy of 11.9~\GeV per nucleon pair. We find that the production of $K^+$ and $K^-$ mesons at mid-rapidity is~$(18.4\pm 6.1)\%$ higher than that of the neutral $K$ mesons.
Although with large uncertainties, earlier data on nucleus-nucleus collisions in the collision center-of-mass energy range $2.6 < \snn < 200$~\GeV are consistent with the present result.
Using well-established models for hadron production, we demonstrate 
that known isospin-symmetry breaking effects and the initial nuclei containing more neutrons than protons lead only to a small (few percent) deviation of the charged-to-neutral kaon ratio from unity at high energies. Thus, they cannot explain the measurements. The significance of the flavor-symmetry violation beyond the known effects is 4.7$\sigma$ when the compilation of world data with uncertainties quoted by the experiments is used.
New systematic, high-precision measurements and theoretical efforts are needed to establish
the origin of the observed large isospin-symmetry breaking.

{\normalfont\bfseries}}

\begin{document}

\maketitle
\pagebreak


\section{Introduction}
One of the main aims of basic research is to understand the fundamental constituents of matter 
and the interactions between them. 
Within Quantum Chromodynamics (QCD)~\cite{Fritzsch:1973pi}, the theory of strong interactions, the fundamental particles are quarks and gluons carrying color -- the charge of strong interactions.
Because of confinement, quarks and gluons are hidden in colorless hadrons, particularly protons and neutrons. The strong force binds them, forming atomic nuclei.

Accelerator-based experiments recording collisions of highly energetic hadrons and nuclei allow for systematic studies of the properties of strong interactions.
In these collisions, many new particles are produced. 
They are predominantly mesons containing one valence quark ($\mathit{q}$) and one valence anti-quark~($\mathit{\overline{q}}$). The most copiously produced are the lightest mesons, pions and kaons, built from {up} ($\mathit{u}$), {down} ($\mathit{d}$) and {strange} ($\mathit{s}$) quarks and the corresponding anti-quarks.

QCD assumes that interactions are independent of quark type (flavor) for equal quark masses and in the absence of other interactions, a feature known as flavor symmetry. When only the light quarks {up} and {down} are considered, flavor symmetry reduces to isospin symmetry, historically introduced in the pre-QCD period by Heisenberg to understand the properties of nuclei~\cite{Heisenberg:1932dw}. 
The masses of {up} and {down} quarks, $m_\mathit{u}=2.16 \pm 0.07$~\MeV and $m_\mathit{d}=4.70 \pm 0.07$~\MeV~\cite{ParticleDataGroup:2024}, are not equal, but they are much smaller than the QCD scale, $\Lambda_{QCD}$ \cite{Kneur:2011vi,Deur:2016tte}. (Note that units in this paper follow the Particle Data Group (PDG)~\cite{ParticleDataGroup:2024} convention: masses and energies are expressed in \MeV (or \GeV), whereas momenta in \MeVc (or \GeVc). The relative differences are given as the ratio of the difference to the mean.) Hence isospin-symmetry breaking effects are small, as confirmed by the mass ratios of pions and kaons, $\left(  m_{\pi^{+}}-m_{\pi^{0}}\right)  /\left(m_{\pi^{+}}+m_{\pi^{0}}\right)  \simeq0.017$
and $\left(  m_{K^{+}}-m_{K^{0}}\right)  /\left(m_{K^{+}}+m_{K^{0}}\right)  \simeq - 0.004$.
Moreover, the elastic cross sections for pion-pion, pion-nucleon, and nucleon-nucleon
scattering closely follow the predictions of isospin symmetry~\cite{Pennington:2005be,Alarcon:2011zs}. 
Here, of special interest is a specific isospin transformation, an inversion of the third component of the isospin, called the charge transformation for historical reasons. It is equivalent to
swapping $u \leftrightarrow d$ quarks. At the hadronic level, the charge transformation implies swapping $p \leftrightarrow n$, $\pi^+ \leftrightarrow \pi^-$, $K^+ \leftrightarrow K^0$, $\overline{K}^{\,0} \leftrightarrow K^-$, etc. 

Let us consider nucleus-nucleus ($A$+$A$) collisions and, for simplicity, assume that both nuclei have an equal number of protons and neutrons.
Without referring to a detailed mathematical formalism, charge symmetry means that strong interactions are invariant under the charge transformation of every nucleus and hadron of the initial and final states. 
For an ensemble of initial states being invariant under the charge transformation, the probabilities of having initial states related by this transformation are equal. This is indeed the case of nucleus-nucleus collisions, for which each nucleus has an equal number of protons and neutrons.
Then, the invariance under charge transformation also holds for the final state ensemble, implying that the mean multiplicities of charge-transformation related hadrons, such as $K^+$ and $K^0$ as well as $\overline{K}^{\,0}$ and $K^-$, coincide: 
\begin{equation}
\langle K^{+} \rangle = \langle K^{0} \rangle \text{ and  } \langle K^{-} \rangle = \langle \overline{K}^{\,0} \rangle 
\text{ . }
\label{eq:kplusminus}
\end{equation}
Note that since models predict only properties of ensembles of events and not outcomes of single events, we need to consider quantities averaged over the event ensembles.
The subject has a vast literature; see, for example
Refs.~\cite{Shmushkevich:55,Shmushkevich:56,MacFarlane:1965wp,Wohl:publ,Pal:2014,Gazdzicki:1991ih}.
Consequently, the exact isospin symmetry prediction for the charged-to-neutral kaon ratio in nucleus-nucleus collisions with electric charge to baryon number $Q/B = 1/2$ reads
\begin{equation}
\label{rk}
R_K\equiv\frac{\langle K^{+} \rangle  + \langle K^{-} \rangle}{\langle K^0 \rangle  + \langle \overline{K}^{\,0} \rangle} = \frac{\langle K^+ \rangle  + \langle K^- \rangle}{2 \langle K_S^{0} \rangle } = 1~.
\end{equation}
(Note that the $\mathit{K}^{\,0}$ and $\mathit{\overline{K}}^{\,0}$ states are produced in strong interactions, but they decay through weak interactions. Consequently, in the final state, one observes linear combinations of the latter known as $\mathit{K}^{0}$ short ($\mathit{K}^{0}_S$) and $\mathit{K}^{0}$ long ($\mathit{K}^{0}_L$), where short and long refer to their weak decay lifetime \cite{ParticleDataGroup:2024}. By neglecting the small CP violation, the multiplicities corresponding to weak and strong eigenstates are related by $\langle K_{S}^{0} \rangle=\frac{1}{2} \langle K^{0} \rangle +\frac{1}{2} \langle \overline{K}^{\,0} \rangle = \langle K_{L}^{0} \rangle$.)
The prediction given by Eq.~(\ref{rk}) is a reference for experimental testing of the isospin symmetry in hadron production processes. For a more detailed introduction and didactic derivations 
see Ref.~\cite{Brylinski:2023nrb}.

Here, we report a measurement of the ratio $R_{K}$ in the 10\% most central collisions of argon ($\mathrm{Ar}$) and scandium ($\mathrm{Sc}$) nuclei at center-of-mass energy per nucleon pair equal to $\snn$ $=$ 11.9~\GeV. 
Further on, we compare the \NASixtyOne result with the world data on charged and neutral kaon production in nucleus-nucleus collisions. The results indicate a significant excess of charged over neutral kaon production. 
This excess cannot be explained by known effects violating the isospin symmetry. This is discussed and demonstrated by comparing experimental results to well-known theoretical approaches, the statistical Hadron Resonance Gas (HRG)~\cite{Vovchenko:2019pjl} and the dynamical Ultrarelativistic Quantum Molecular Dynamics (UrQMD)~\cite{Bleicher:1999xi} models. The predictions of models are calculated for reactions corresponding to experimental data, generally with $Q/B < 1/2$. They consider isospin-breaking effects in strong interactions and, importantly, the production and subsequent decays of the $\phi$ mesons.

Summarizing, the \NASixtyOne Collaboration measures a charged-to-neutral kaon ratio $R_K = 1.184 \pm 0.061$ in Ar+Sc collisions at 11.9 \GeV per nucleon pair. This value aligns with previous experimental measurements, albeit their uncertainties are larger. 
The significance of the isospin symmetry violation beyond the known effects amounts to 4.7$\sigma$ 
when all measurements are considered, and uncertainties quoted by the experiments are used. This is the first evidence of an unexplained isospin symmetry violation in hadron production processes.

\section{Results}
\subsection{\texorpdfstring{Production of $\pmb{K}$ mesons}{Production of K mesons} in central Ar+Sc collisions at the CERN SPS}
The new experimental results presented here have been obtained by the \NASixtyOne fixed-target experiment at the CERN Super Proton Synchrotron~\cite{fac_paper}. The measurements of $\mathit{K^+}$ and $\mathit{K^-}$ production in the 10\% most central Ar+Sc reactions at $\snn=11.9$~\GeV have been published elsewhere~\cite{NA61SHINE:2023epu}. 
The analysis procedure and details of systematic uncertainties are given in 
Refs.~\cite{Lewicki:2772291,Podlaski:2799198}.
Here, we present the first measurement of $\mathit{K}^{0}_S$ production in nucleus-nucleus collisions from \NASixtyOne. Earlier data from this experiment, on $\mathit{K}^{0}_S$ production in $p$+C, $\pi^-$+C, $\pi^+$+C, $\pi^+$+Be, and $p$+$p$ collisions can be found in Refs.~\cite{ 
NA61SHINE:2015bad_k0s_k+k-_p+C_31GeV,
NA61SHINE:2022uxp_Brant_k0s_pC_120GeV,
NA61SHINE:2022tiz_Michael_k0s_k+k-_pi-+C,
NA61SHINE:2019nzr_k0s_k+k-_pi+C_pi+Be,
NA61SHINE:2021iay_k0s_pp, NA61SHINE:2024eqz}.  
For more details concerning the experimental procedure, see Methods' "Experimental procedure" subsection.

\begin{figure}
   \centering   
   \includegraphics[width=.87\linewidth]{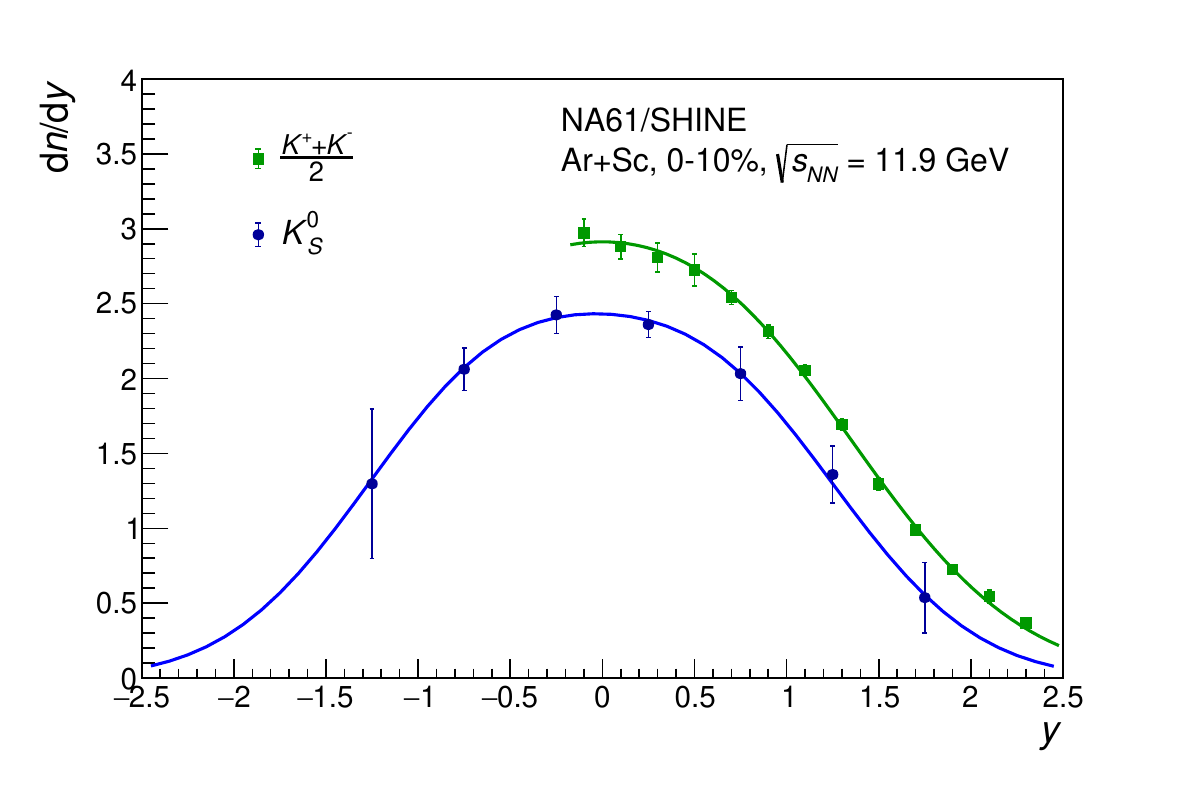}
   \caption{{\normalfont\bfseries Comparison of rapidity spectrum of neutral ($\pmb{K^0_S}$) with the averaged spectrum of charged ($\pmb{K^+}$ and $\pmb{K^-}$) mesons} in the 10\% most central Ar+Sc collisions at $\snn=11.9$~\GeV.
     Total uncertainties, calculated as the square root of the sum of squared statistical and systematic uncertainties ($\sqrt{\sigma_{stat}^2 + \sigma_{sys}^2}$) are drawn. For charged $K$ mesons, the total uncertainties were calculated separately for positively and negatively charged and then propagated.}
    \label{fig:y_75comp}
\end{figure}

The comparison of the rapidity distribution of $\mathit{K}^{0}_S$ mesons to the average of rapidity distributions for $\mathit{K^+}$ and $\mathit{K^-}$ mesons is presented in Fig.~\ref{fig:y_75comp}. The rapidity $y$ is a relativistic generalization of the particle velocity along the direction of the incoming nuclei. We calculate rapidity in the nucleon-nucleon collision center-of-mass system, and positive $y$ corresponds to the direction of the Ar nucleus. 

In the entire range of rapidity covered by the measurement, the averaged charged $\mathit{K}$ mesons yield prevails significantly over the neutral $\mathit{K}^{0}_S$ mesons one. To quantify this effect, Table~\ref{tab1} presents the rapidity densities $\mathrm{d}n/\mathrm{d}\y$ of $\mathit{K^+}$, $\mathit{K^-}$ and $\mathit{K}^{0}_S$ production measured at mid-rapidity ($y\approx 0$). Here, the relative excess of charged mesons is $\mathrm{(18.4 \pm 6.1)}$\%. Integration of the two distributions in Fig.~\ref{fig:y_75comp} over positive rapidity, $y>0$, gives $4.28 \pm 0.13$ and $3.22 \pm 0.37$ for the production rates per collision of $\mathit{(K^++K^-)/2}$ and $\mathit{K}^{0}_S$, respectively (total uncertainties are given; the quantities provided for charged $\mathit{K}$ mesons are based on Ref.~\cite{NA61SHINE:2023epu}). The resulting difference of $1.06 \pm 0.39$ corresponds to a surplus of charged ($\mathit{K^+}$ and $\mathit{K^-}$) over neutral ($\mathit{K}^{0}$ and $\mathit{\overline{K}}^{\,0}$) states equal to $2.12 \pm 0.79$ at positive rapidity. Under the assumption that the charged-to-neutral ratio would be similar also at negative rapidity, the total excess would amount to $4.2 \pm 1.6$ additional $K^+$ or $K^-$ mesons per one central Ar+Sc collision.

\begin{table}
\vspace*{0.2cm}
 \begin{center}
\begin{tabular}{@{}lllllll@{}}
\toprule
&                            &      & statistical  & systematic &  ~~~~total~~~~  &\\
&$\left(\frac{\mathrm{d}n}{\mathrm{d}y}\right)_{\hspace*{-0.1mm}y\hspace*{0.3mm}\approx\hspace*{0.3mm}0}\mathit{(K^+)}$
    &3.732& $\pm$ 0.016  & $\pm$ 0.148 & $\pm$ 0.149 &\\
&$\left(\frac{\mathrm{d}n}{\mathrm{d}y}\right)_{\hspace*{-0.1mm}y\hspace*{0.3mm}\approx\hspace*{0.3mm}0}\mathit{(K^-)}$
    &2.029& $\pm$ 0.012  & $\pm$ 0.069 & $\pm$ 0.070  &\\
&$\left(\frac{\mathrm{d}n}{\mathrm{d}y}\right)_{\hspace*{-0.1mm}y\hspace*{0.3mm}\approx\hspace*{0.3mm}0}\mathit{(K}^{0}_S{\mathit)}$ 
      &2.433& $\pm$ 0.027  & $\pm$ 0.102 & $\pm$ 0.106  &\\
\midrule
&charged-to-neutral $\mathit{K}$ meson ratio:& &  & &\\
~~~~~~~~
&${R_K}$
    & 1.184 & $\pm$ 0.014 & $\pm$ 0.060 & $\pm$ 0.061   &~~~~~~~\\
\bottomrule
\end{tabular}
\end{center}
\caption{{\normalfont\bfseries Rapidity densities of charged and neutral $\pmb{K}$ mesons produced at mid-rapidity.} 
  The measurement was performed in the 10\% most central Ar+Sc collisions at $\snn=11.9$~\GeV, as described in Methods' "Experimental procedure" subsection. 
  The excess of charged over neutral mesons is quantified by the ratio ${R_K}$ defined in Eq.~(\ref{rk}).}\label{tab1}%
\end{table}

\begin{figure}[h!]
   \centering   
   \hspace*{-0.1cm}\includegraphics[width=.9\linewidth]{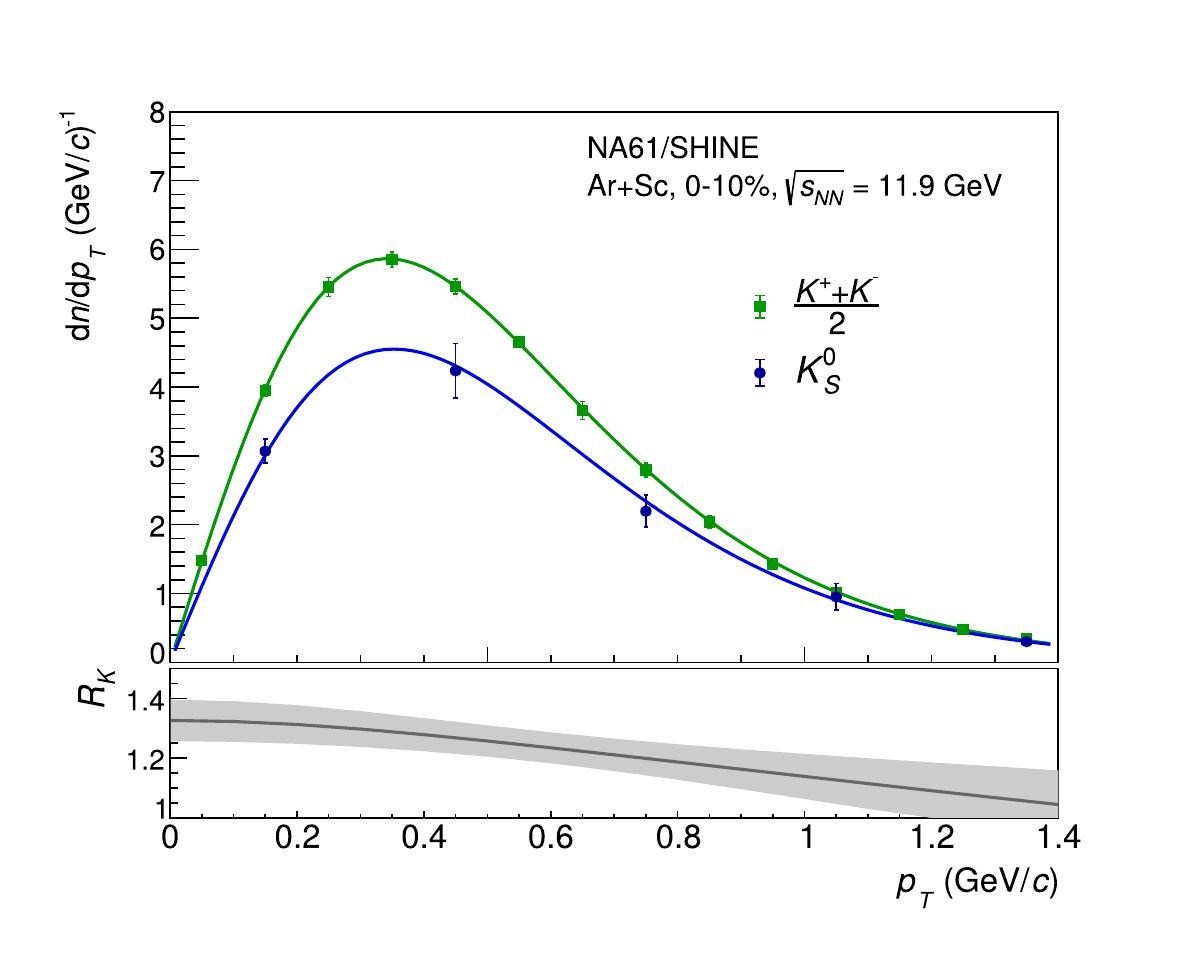}
   \vspace*{-0.2cm}
   \caption{{\normalfont\bfseries Comparison of transverse momentum spectrum of neutral ($\pmb{K^0_S}$) with the averaged spectrum of charged ($\pmb{K^+}$ and $\pmb{K^-}$) mesons} in the 10\% most central Ar+Sc collisions at $\snn=11.9$~\GeV.
     The bottom panel shows the ratio of the two distributions, as defined in Eq.~(\ref{rk}).
     The meaning of the total uncertainties drawn is the same as in Fig.~\ref{fig:y_75comp}.}
    \label{fig:y_75comppt}
\end{figure}

A comparison of distributions of $\mathit{K}^{0}_S$ with averaged $\mathit{K^+}$ and $\mathit{K^-}$ mesons as a function of transverse momentum $p_{T}$ (the momentum component perpendicular to the direction of the incoming nuclei) is shown in Fig.~\ref{fig:y_75comppt}. Both distributions are integrated over the rapidity range $0<y<2$. The prevalence of charged over neutral $\mathit{K}$ mesons is again evident. The insert in the figure shows the $p_{T}$-dependence of the ratio ${R_K}$. The corresponding excess of $\mathit{K}$ mesons containing $\mathit{u}$, $\mathit{\overline{u}}$ over those containing $\mathit{d}$, $\mathit{\overline{d}}$ quarks and anti-quarks remains in the range 6--33\% over the considered range of $p_{T}$.

\newpage

\subsection{Comparison to the world data and models}

\begin{figure}[h]
\centering
\includegraphics[width=\textwidth]{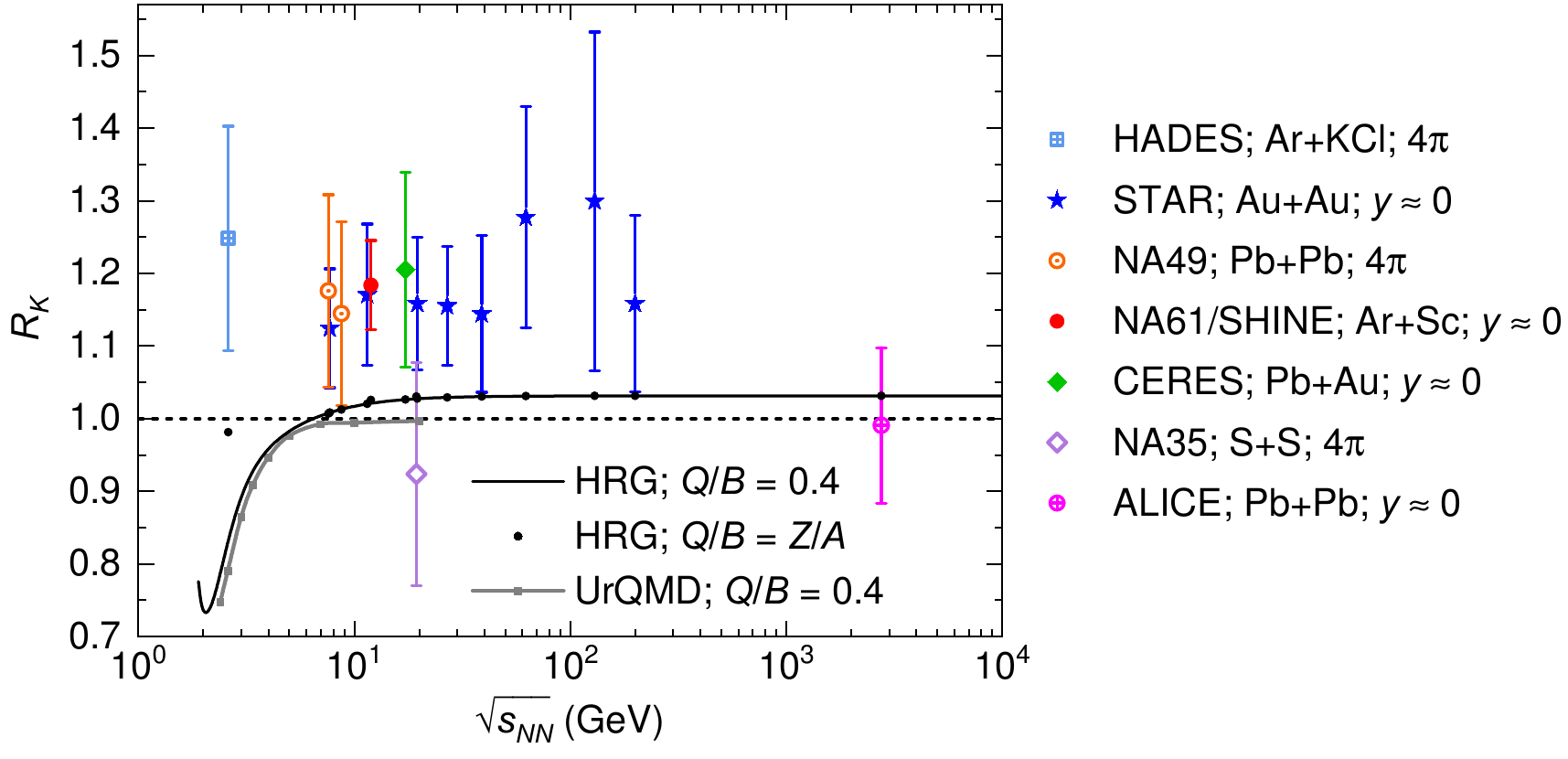}
\caption{\label{fig:kaon-ratio} 
{\normalfont\bfseries 
The charged-to-neutral kaon ratio $\pmb{R_K}$ as a function of collision energy. 
}
The symbols show the experimental world data with total uncertainties; see Methods' "World data" subsection for details. The black line shows the HRG predictions for $Q/B=0.4$.
The black dots indicate the HRG predictions for $Q/B$ values corresponding to the ones in the experiments. For different nuclei, $Q/B$ corresponds to the electric charge over the baryon number of the whole system. The gray squares show UrQMD predictions.
See Methods' "Models" subsection for details on models. }
\end{figure}

Figure~\ref{fig:kaon-ratio} compares the present measurement of the ratio ${R_K}$ at mid-rapidity and the world data compiled by us and detailed in Methods' "World data" subsection.
The experimental results were obtained by
CERES~\cite{Kalisky:2008ozf, Radomski:2008zz, RadomskiPhD}, 
STAR BES~\cite{STAR:2017sal, STAR:2019bjj}, 
STAR~\cite{STAR:2002hpr, STAR:2008med, STAR:2010yyv, STAR:2011fbd}, 
ALICE~\cite{ALICE:2013mez, ALICE:2013cdo}, 
NA35~\cite{NA35:1992njd, NA35:1994veo}, 
NA49~\cite{NA49:2007stj, NA49:2002pzu, Strabel:phd}, 
and HADES~\cite{HADES:2009lnd, Agakishiev:2010zw} 
experiments.
We note that the compilation includes measurements at mid-rapidity and total multiplicities. This may increase the overall spread between the data points. We also note the sizeable uncertainties of the earlier measurements. These probably explain why the aforementioned charged-over-neutral anomaly was never reported as an experimental observation. Despite these uncertainties, a consistent picture emerges in the energy range $2.6 < \snn < 200$~\GeV. The ratio is above one for all experiments except NA35 and ALICE.

\begin{figure}
\centering
\includegraphics[width=0.8\textwidth]{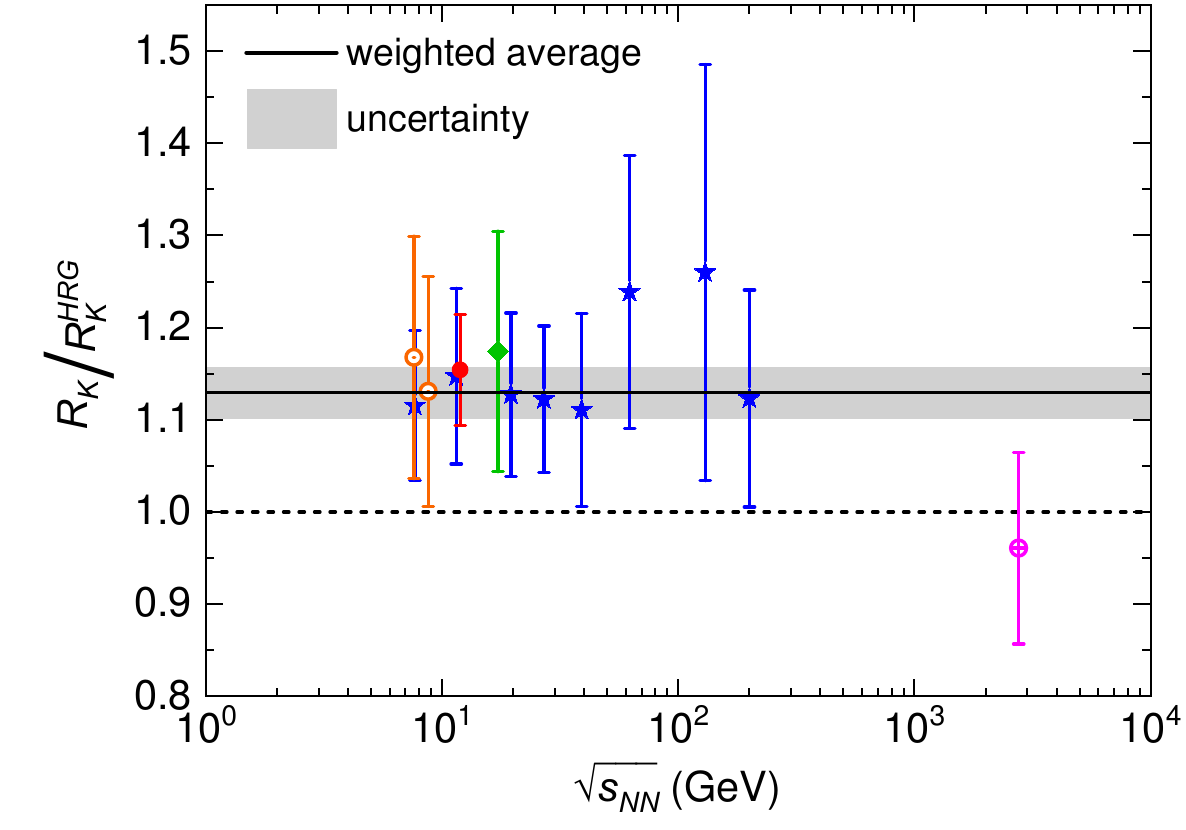}
\caption{\label{fig:data-vs-hrg} 
{\normalfont\bfseries
The experimental data for the charged-to-neutral kaon ratio divided by the HRG baseline $\pmb{R_K/R_K^{HRG}}$ as a function of collision energy. 
}
The symbols are explained in Fig.~\ref{fig:kaon-ratio}.
The solid black line shows the weighted average of the experimental data, and the shaded area shows the uncertainty of the weighted average. }
\end{figure}

Figure~\ref{fig:kaon-ratio} also compares the data with the HRG and UrQMD model predictions.
The HRG calculations were performed with $Q/B = 0.4$ (solid line) and $Q/B$ values corresponding to collisions studied in the experiments (black dots). The UrQMD results were obtained for central Au+Au collisions. See Methods' "Models" subsection for details on the models. The predictions of models agree well with each other but
are systematically lower than the experimental data.
At energies larger than 10~\GeV, the mass difference between charged and neutral kaons, leading to isospin-symmetry breaking (mostly via $\phi$-meson decays), increases 
the $R_K$ ratio by about 0.03. Other isospin-breaking effects can be neglected. The ratio $R_K$ is reduced for collisions with $Q/B < 1/2$, but this is insignificant for energies larger than 10~\GeV. At energies lower than 10~\GeV, the $R_K$ ratio is significantly more sensitive to the known isospin-breaking effects as well as to the $Q/B$ ratio; see Methods' "Models" subsection for more details.

To quantify the measured isospin-symmetry breaking beyond the known effects, the ratio of the measured $R_K$ to the corresponding HRG baseline is shown in Fig.~\ref{fig:data-vs-hrg}.
We do not consider the lowest energy data point (HADES) because the known isospin-breaking and $Q/B < 1/2$ effects are significant at the low collision energies; thus, predictions may be model-dependent. 
We also do not consider the NA35 point because, unlike other measurements, charged kaons were identified by reconstructing their decays, leading to large statistical uncertainties and possible biases. Thus, the number of selected measurements at different collision energies (from SPS to LHC) is $13$.
Out of them, only one is below unity.
No significant dependence of the double ratio on collision energy and nuclear mass number of colliding nuclei is visible.

The weighted average of all double ratios shown in Fig.~\ref{fig:data-vs-hrg} is $1.129 \pm 0.027$,
where the uncertainty was calculated using kaon uncertainties reported by experiments. The HRG uncertainties are small and were neglected. The significance of the isospin violation is $4.7 \sigma$. 
The $\chi_{\textit min}^2/\text{dof} \approx 0.3$
may indicate either a correlation between results or an overestimation of the uncertainties.

\section{Discussion}

In the following, we discuss possible effects that may potentially contribute to the violation of isospin symmetry in kaon production.

First, we consider symmetry-breaking effects due to the non-equal bare $u$ and $d$ masses in strong interactions. They are included in the HRG and UrQMD models. Then, we discuss the possible influence related to electromagnetic and weak processes.

\begin{enumerate}[(A)]
\item 
\vspace{0.1cm}
Mass effects within strong interactions.
Within QCD, the isospin symmetry is not exact because $u$ and $d$ quark masses are different, $\approx 2.2$ and $\approx 4.7$~\MeV, respectively.
The different quark masses lead to different masses of hadrons within the isospin multiplets, particularly different masses of charged and neutral kaons. This modifies the ratio $R_K$.
We list below the considered effects and quantify their influence on $R_K$
using the statistical Hadron Resonance Gas model (HRG)~\cite{Vovchenko:2019pjl}.
The results are cross-checked with the microscopic
transport model, UrQMD~\mbox{\cite{Bass:1998ca,Bleicher:1999xi,Bleicher:2022kcu}}.
For details, see Methods' "Models" subsection.

\vspace{0.1cm}
\begin{enumerate}[(i)]
    \vspace{0.1cm}
    \item 
    Smaller masses of charged kaons than neutral ones, $m_{K^+}=m_{K^-}=493.7$~\MeV and $m_{K^0}=m_{\overline{K}^{\,0}}=497.6$~\MeV,
    lead to an increase of $R_K$ resulting from direct kaon production by about 0.02. 
    This was estimated by removing resonances from the particle list of HRG. We have numerically verified that HRG for $Q/B =1/2$ and with exact isospin symmetry gives $R_K = 1$, as expected. 
    \vspace{0.1cm}
    \item 
    A significant fraction of kaons (up to several dozen percent) results from resonance decays~\cite{Becattini:2005xt}. Different kaon masses affect the branching ratios 
    of resonances. 
    The most striking example is $\phi(1020)$ meson, which decays about $1.45$ more frequently into charged kaons than neutral ones. This large difference is because the $\phi(1020)$ mass is just above the kaon-kaon thresholds. Including the kaon production from resonance decays
    increases $R_K$ by about 0.03.

 \vspace{0.1cm}
    \item 
In connection to the previous point, other potentially relevant $K\overline{K}$ decays refer to the resonances $a_{0}(980)$ and $f_{0}(980)$, whose masses are close to the $K\overline{K}$ decay threshold. 
In the PDG review~\cite{ParticleDataGroup:2024}, for both resonances $K\overline{K}$ decays are
reported as seen. The HRG model used in this work 
initially assumes equal branching ratios ($BR$) of 
$K^{+}K^{-}$ and $K^{0}\overline{K}^{\,0}$ decay channels. Yet, just as for the $\phi$
meson, isospin breaking may be relevant. (Note that
isospin breaking for $a_{0}(980)$ and $f_{0}(980)$ is experimentally
confirmed by the $a_{0}(980)$ - $f_{0}(980)$ mixing~\cite%
{Achasov:1979xc,Wu:2008hx} leading to the otherwise forbidden decays $%
a_{0}^{0}(980)\rightarrow f_{0}(980)\rightarrow \pi \pi $ and $%
f_{0}(980)\rightarrow a_{0}^{0}(980)\rightarrow \pi \eta $; see the
experimental results for these small but non-zero transitions in Ref.~\cite%
{BESIII:2018ozj}.) Using the estimate (see Methods' "Models" subsection) $BR(K^{+}K^{-})$/$BR(K^{0}\overline{K}
^{\,0})\approx 1.2$ one gets an increase of $R_{K}$ by about $0.5\%$ . 
The yield of charged kaons may also be
affected by other effects, such as the  electromagnetic interaction
between $K^{+}$ and $K^{-}$ (see also (B)~(ii) below). 
To calculate the upper limit due to this effect, we assumed $BR(K^{+}K^{-})$ $=$ $BR(K\overline{K})$ (no decays to neutral kaons) for both $a_{0}^{0}(980)$ and $f_{0}(980)$.
This leads to the increase of $R_{K}$
by at most $3\%$ at the highest collision energy.

    \vspace{0.1cm}
    \item  
    Mass differences of hadrons from other isospin multiplets also break the flavor symmetry and affect $R_K$. 
    The largest effect comes from the mass difference between proton and neutron, which reduces $R_K$ at the lowest collision energies.
    At energies larger than 10~\GeV, this effect is negligible.
\end{enumerate}
\vspace{0.1cm}

\vspace{0.1cm}
\item 
Electromagnetic processes. Electromagnetic interactions do not obey isospin symmetry because electric charges differ
for the quark flavors $u$ and $d$. The electromagnetic interaction slightly affects the masses of hadrons. For instance, the neutral pion is lighter than the charged ones. The HRG and UrQMD include such effects since the physical masses are used.  However, the effects mentioned below are not included in the models.

\begin{enumerate}[(i)]
\item
 Electromagnetic decays of hadrons are typically suppressed by a factor $\alpha \simeq 1/137$ compared to strong ones. Consequently, decays that involve the production of virtual photons and their subsequent decay into charged kaons are suppressed by a factor $\alpha^2$ and thus negligible. Taking into account the charge of the nuclei $Z_1$ and $Z_2$, one would expect an effect of the type $Z_1Z_2\alpha^2$, which is not observed in the experimental data -- isospin-symmetry breaking for collisions of light and heavy nuclei is similar, see Fig.~\ref{fig:data-vs-hrg}.

 \item 
There may also be non-perturbative electromagnetic
effects at the hadronic level that might affect the kaon multiplicities. One
notable example is the case of $K^{+}K^{-}$ pairs with low momenta of kaons
$p \lesssim m_{K}\alpha\simeq3.6$ \MeVc, see e.g. Ref.~\cite{Krewald:2003ab}. This is,
in particular, the case for the decays of the resonances $a_{0}(980)$ and
$f_{0}(980)$, whose masses are close to the two-kaon decay
threshold. It is then possible that this effect will increase the number
of produced $K^{+}K^{-}$ pairs.  Experimental search for $K^{+}%
K^{-}$ interaction close to the threshold was conducted at COSY in the study of
the reaction $p+p\rightarrow p+p+K^{+}+K^{-}$, see e.g. Ref.~\cite{Ye:2012ae} and
the review~\cite{Wilkin:2016mfn}. An increase in the cross-section
close to the threshold has been measured, where different effects
appear to be relevant, with an important role played by the $pK^{-}$ Coulomb
attraction.  We have quantified the impact of
$a_{0}(980)$ and $f_{0}(980)$ decays
in point (A)~(iii) above and in Methods' "Models" subsection, showing that they cannot explain the measured value of $R_K$.

 \item 
The $u\overline{u}$ and $d\overline{d}$ pair creation in strong processes may be affected by electromagnetic interactions. They are different for $u\overline{u}$ and $d\overline{d}$ pairs due to different electric charges of up and down quarks. This leads to a different phase space for their production, favoring $u\overline{u}$ pairs and thus charged kaons. In particular, the quark-gluon effective coupling is enhanced by QED effects due to the attraction among quarks, leading to a larger coupling of gluons to $u$-quarks than $d$-quarks~\cite{Goldman:1989as}. 
A model of the space-time evolution of the pair creation will be needed to quantify the effect.
In addition, the isospin breaking due to the Coulomb potential of highly-charged fireballs formed in heavy-ion collisions is discussed in Ref.~\cite{Letessier:1998ca} within the statistical QGP model. 
We recall that electromagnetic interactions are expected to modify fusion rates in
the Big Bang nucleosynthesis epoch; see, for example, 
Refs.~\cite{Salpeter:1954nc,Grayson:2024uwg}.
 \end{enumerate}

\vspace{0.1cm}
\item 
Uncertainties in weak decays. The weak interaction does not obey the isospin symmetry.
The mean lifetimes~\cite{ParticleDataGroup:2024} of charged and neutral kaons are
$\tau(K^+) = \tau(K^-) = (1.2380 \pm 0.0020) \cdot 10^{-8}$~s
and
$ \tau(K^0_S) = (8.954 \pm 0.004) \cdot 10^{-11}$~s,
$ \tau(K^0_L) = (5.116 \pm 0.021) \cdot 10^{-8}$~s.
The charged kaons are typically measured by reconstructing their trajectories in a detector. Due to the large mean lifetime, the corrections for their losses caused by weak decays are small. 
In contrast, 
the neutral $K^0_S$ kaons are measured by reconstructing their decays into two charged pions. Typically, the corrections for the losses caused by weak decays are large. This is because the decay should be far enough from the interaction point to separate the decay point from the background in high-multiplicity $A$+$A$ collisions.
Assuming that the $K^0_S$ meson is measurable when the lifetime of a particle in its rest frame
is larger than the mean lifetime (typical for \NASixtyOne), one estimates the maximum relative bias of the mean multiplicity, $\Delta(\langle K_S^{0} \rangle)/\langle K_S^{0} \rangle$, as:
\begin{equation}
\label{bias}
\frac
{\Delta(\langle K_S^{0} \rangle)}  {\langle K_S^{0} \rangle}
= 
\frac
{3 \cdot \sigma(\tau(K_S^{0}))}  {\tau(K_S^{0})} \approx 0.0013\, ,
\end{equation}
where $\sigma(\tau(K_S^{0}))$ is the uncertainty of the mean $K^0_S$ lifetime.
Thus, the maximum deviation of $R_K$ from unity due to the uncertainty of the mean $K^0_S$ lifetime is 0.13\%. 
\end{enumerate}

Finally, we discuss the consequences of having collisions with \boldmath{$Q/B < 1/2$}, which corresponds to many experimental results presented in Fig. \ref{fig:kaon-ratio}. The third component of isospin equals $|I_z| = |B/2 - Q|$ and therefore the total isospin is limited as
$|I_z| \leq I \leq B/2$. The compiled experimental results in nucleus-nucleus collisions correspond
to the $Q/B$ ranging from about 0.4 (Pb+Pb and Au+Au collisions) to 0.5 (S+S collisions); see Fig.~\ref{fig:kaon-ratio}. 
These limits correspond to the normalized per baryon third component
and total isospin $|I_z|/B = 0.1$, $0.1 < I/B <1/2$ and
$|I_z|/B = I/B = 0$, respectively. The non-zero $I_z$ and $I$
for heavier nuclei can affect the charged-to-neutral kaon ratio in two ways:
\vspace{0.1cm}
\begin{enumerate}[(a)]
    \vspace{0.1cm}
    \item The larger fraction of neutrons than protons for heavy nuclei having $|I_z|/B \approx 0.1$ enhances neutral kaon production
    compared to charged ones and thus reduces $R_K$. 
    This fact is taken into account by the employed theoretical models that use the physical value for $Q/B$.
    The reduction of $R_K$ is significant at low collision energies and is small compared to other effects at energies larger than 10~\GeV; see Fig.~\ref{fig:kaon-ratio} and Methods' "Models" subsection.  
    \vspace{0.1cm}
    \item  
    For $|I_z|/B \approx 0.1$, the total isospin is limited by
    $0.1 \leq I/B \leq 1/2$. Generally, nuclei in the ground state have the lowest possible value of the total isospin~\cite{Lenzi2009}.
    This rule extends to a state of two identical nuclei in the ground state, which, for the considered case, implies $I/B \approx |I_z|/B \approx 0.1$. 
    Thus, a possible dependence of $R_K$ on $I$ and $I_z$ reduces to the dependence on $I = |I_z|$. The latter is discussed above in point (a).
    The experimental data also allow a rough estimate of the influence of the small but non-zero value of the normalized isospin on the charged-to-neutral kaon ratio. 
    Results for heavy nuclei, $^{208}$Pb+$^{208}$Pb and 
    $^{197}$Au+$^{197}$Au ($R_K \approx 1.15$ for $I/B = |I_z|/B \approx 0.1$), are similar  to those 
    for intermediate nuclei, $^{40}$Ar+$^{45}$Sc ($I/B = |I_z|/B \approx 0.04$), and $^{32}$S+$^{32}$S ($I/B = |I_z|/B = 0$), see Fig.~\ref{fig:kaon-ratio}.
    This suggests that the sensitivity of $R_K$ to the $I/B = |I_z|/B$ close to zero ($ I/B = |I_z|/B \leq 0.1$) is low.
    Note, however, that there are large uncertainties in the experimental results.

    \vspace{0.1cm}
    \item 
    In the case of central collisions of heavy nuclei, the charge-to-baryon ratio of the interacting nucleon system (participant nucleons) can be higher than the total proton-to-nucleon ratio in colliding nuclei. This is because protons tend to be distributed closer to the center of a nucleus~\cite{Thiel:2019tkm}. However, this effect is expected to be small because the ratio $R_K$ seems independent of the colliding nuclei size. Low-mass nuclei can be described as alpha clusters~\cite{Otsuka:2022bcf} (clusters of two protons and two neutrons). Thus, if significant, the effect should disappear for collisions of low-mass nuclei, particularly at the lowest collision energy ($\lesssim 4$~\GeV), where $R_K$ is sensitive to the small changes of the electric charge to baryon number ratio, $Q/B$.     
    The data do not support this. 
\end{enumerate}

\vspace{0.2cm}

Closing comments on future perspectives of experimental and theoretical efforts are in order here.

\begin{enumerate}[(I)]
    \vspace{0.1cm}
    \item
    Concerning measurements, reviewing the validity of the past results and confirming them with new high-precision data is important.
    Systematic results on the collision energy and nuclear mass dependence of the isospin-breaking effect should help us understand its nature.
    Measurements of the charged-to-neutral kaon ratio in collisions of an equal
    number of protons and neutrons would reduce uncertainty in its interpretation. The \NASixtyOne experiment plans to perform such measurements for O+O and Mg+Mg collisions~\cite{Mackowiak-Pawlowska:2867952}. If needed, the measurements of deuteron-deuteron interactions may be possible in the long future. They will require the production of primary deuteron beams in the CERN accelerator complex and the construction of a liquid deuterium target for \NASixtyOne.
    
    \vspace{0.1cm}
    \item
    In connection with the previous point, an interesting experimental test is also possible by considering $\pi^-$+C and $\pi^+$+C interactions~\cite{Kowalski:2907307}. While the ensemble of only one of them is not invariant under charge transformation, the ensemble having an equal number of $\pi^-$+C and $\pi^+$+C interactions is invariant. Thus, for the joint ensemble, the exact charge symmetry predicts $R_K = 1$. \NASixtyOne recorded data on $\pi^-$+C and $\pi^+$+C interactions at 158~\GeVc in October 2024 for the test.
    
    \vspace{0.1cm}
    \item
    During the current CERN accelerator complex operation period (Run~3), \NASixtyOne records high-statistics data on inelastic Pb+Pb collisions at 150\AGeVc. This will allow precision measurements of the $R_K$ ratio as a function of collision centrality. The larger number of neutrons than protons in Pb nuclei may reduce the ratio $R_K$. This should be taken into account when interpreting the results.

    \vspace{0.1cm}
    \item
    The relationship between the production rates of charged and neutral kaons in hadronic collisions was used to estimate the flux of secondary neutral kaons produced in $p$+Be collisions at beam momenta of several 100~\GeVc in neutral kaon and neutrino beams~\cite{Bonesini:2001iz}.  Recently, the charged-to-neutral kaon ratio was studied in $p$+$p$ interactions at SPS energies, and no excess in comparison to a simple quark counting model emerged~\cite{Stepaniak:2023pvo}.

    \vspace{0.1cm}
    \item
       Kaons play a special role due to their simple isospin structure and easy measurement. This explains why the first results on a large isospin-symmetry breaking in multi-particle production are reported for kaons. Yet, it is important to perform a similar study for other isospin multiplets in the future.
       For example, using the same methods, we have checked that the (anti-)proton to (anti-)neutron ratio is even less sensitive to the known isospin-symmetry breaking effects and, thus, is predicted to be almost exactly one in nucleus-nucleus collisions with $Q/B=0.5$ in a broad range of collision energies including the low energies. 

    \vspace{0.1cm} 
    \item 
     One can extend the current models by introducing new isospin-breaking processes and fitting their parameters to the data. 
    This can be done either for the quark-gluon processes or the hadron-resonance processes.
    For example,  within the statistical hadronization models, one can introduce the quark fugacity factors for $u$ and $d$ quarks separately~\cite{Petran:2013lja,Rafelski:2014cqa,Rafelski:2015hta}. This could allow us to make predictions for other hadron ratios but will not explain the origin of the violation.
   
    \vspace{0.1cm}
    \item 
    The possibility of having a phase of strongly interacting matter with a significant isospin violation was suggested by Pisarski and Wilczek within a linear $\sigma$ model of QCD~\cite{Pisarski:1983ms}.
    They expect masses of $\pi^0$ and $\eta$ mesons to decrease if the $U_A(1)$ symmetry is effectively restored at temperatures lower than the one of the chiral phase transition.
    The chiral anomaly~\cite{Giacosa:2017pos,Giacosa:2023fdz} could break isospin (especially, it affects the pion isotriplet)  but does not affect the charged-to-neutral kaon ratio. 
    
    \vspace{0.1cm}
    \item 
    Creating Disoriented-Chiral-Condensate (DCC) domains in heavy-ion collisions has been considered for many years~\cite{Anselm:1991pi,Blaizot:1992at,Rajagopal:1993ah}.
    They may be signaled by large fluctuations of the charged-to-neutral pion~\cite{Bjorken:1993wj} and kaon ratios~\cite{Schaffner-Bielich:1998mra,Gavin:2001uk}. A puzzling result on kaon fluctuations was recently reported by ALICE at LHC~\cite{ALICE:2021fpb}.
    Its possible interpretation by the DCC or disoriented-isospin-condensates formation is discussed in Refs.~\cite{Kapusta:2022ovq,Kapusta:2023xrw}. The considered models for the charged-symmetric ensemble of collisions predict $R_K = 1$~\cite{SQM24}. The inclusion of isospin-breaking effects in the extended Linear Sigma model~\cite{Parganlija:2012fy} was recently discussed in 
    Ref.~\cite{Kovacs:2024cdb}, where through a fit to available experimental masses and decays of light mesons, it is shown that the relative difference between the $u\overline{u}$ and $d\overline{d}$ chiral condensates amounts to $0.02 \% $, implying that only very small deviations from $R_K =1$ are expected from this effect.
       
\end{enumerate}

Thus, the presented results on the charged-to-neutral kaon ratio are the first evidence of an unexplained isospin symmetry violation in hadron production processes.
Further studies are needed to understand the underlying physics, particularly reducing the experimental uncertainties and quantifying the role of electromagnetic effects. If these steps do not solve the issue, more speculative explanations shall be investigated.


\clearpage

\noindent
{\Large\normalfont\bfseries Methods}\\


{\normalfont\bfseries A. Experimental procedure}

\noindent
{\normalfont\bfseries Experimental setup.}
The SPS Heavy Ion and Neutrino Experiment (SHINE) is a fixed-target detector operating at the CERN Super Proton Synchrotron (SPS). It is a multi-purpose spectrometer optimized to study hadron production in various collisions (hadron-proton, hadron-nucleus, and nucleus-nucleus). The detection setup used for the measurements reported here is described below. Its details and a description of the detector performance can be found in Ref.~\cite{fac_paper}.

The beamline is equipped with an array of beam detectors upstream and downstream of the target, used to identify and measure the trajectory of the beam particles and trigger the spectrometer data acquisition. The tracking devices of the \NASixtyOne spectrometer are Time Projection Chambers (TPCs). Two Vertex TPCs are placed inside a magnetic field. Two large-volume Main TPCs measure the charged particle trajectories downstream of the 4.5 Tm magnetic field. The latter provides the bending power for a precise determination of particle momenta. The information about the energy losses (\dedx) of the charged particles in the TPCs, together with Time-of-Flight (ToF) measurements, allows for particle identification in a wide momentum range. The most downstream detector on the beamline is the Projectile Spectator Detector (PSD). It measures the energy of the spectator remnant of the projectile nucleus, closely related to the collision centrality in nucleus-nucleus reactions.

\vspace*{0.2cm}\noindent
{\normalfont\bfseries Physics objects.}
This article compares the production of charged and neutral $\mathit{K}$ mesons in Ar+Sc collisions at a center-of-mass energy per nucleon pair of 11.9~\GeV. The $^{40}_{18}\mathrm{Ar}$ beam had a momentum of 75\AGeVc. The stationary target consisted of six $^{45}_{21}\mathrm{Sc}$ plates, with a total thickness of 6~\mm.

A detailed account on the extraction of charged ($\mathit{K^+}$, $\mathit{K^-}$) yields can be found in Ref.~\cite{NA61SHINE:2023epu}. Only the neutral \Ks mesons are considered in the present analysis.
They can be detected via their weak decay into two charged pions $(\Ks \rightarrow \pip \pim)$. The mean lifetime $(c\tau)$ for this decay is 2.7~cm. A detailed presentation of the \Ks analysis procedure and systematic uncertainties can be found in Ref.~\cite{Brylinski:phd}.

\vspace*{0.2cm}\noindent
{\normalfont\bfseries Analysis.}
Before analyzing \Ks mesons, the recorded Ar+Sc collision data undergo event and track selection procedures. Event selection uses information from the beam detectors to ensure the quality of the measured beam trajectory. It rejects events with more than one beam-target interaction during the trigger-time window. It also reduces the background from off-target interactions based on information about the quality of the main interaction vertex. Finally, it selects the 10\% most central collisions using the information from the PSD. This is realized by selecting the 10\% lowest energy deposits from the spectator remnant of the Ar nucleus. The total number of recorded collisions (events) was $2.77 \cdot 10^6$, from which $1.03 \cdot 10^6$ (37\%) remained after all cuts. For more details, especially the centrality selection, see Ref.~\cite{NA61SHINE:2023epu}. 

The next step is reconstructing the charged particle tracks in the TPCs. Pattern recognition algorithms combine space points recorded in the TPCs into tracks. Their curvature and the magnetic field are used to compute the momenta of the corresponding particles. The minimum number of reconstructed space points in the VTPCs must be more than 10, and the computed momenta must be larger than 400~\MeVc (in the laboratory frame). The latter selection excludes a large fraction of low-momentum electrons from the analysis. The known positions of the target and the most probable intersection point of measured tracks define the position of the primary vertex. 

\vspace*{0.2cm}\noindent
{\normalfont\bfseries $\pmb{K^0_S}$ reconstruction.}
Unlike the charged particles, the neutral $\mathit{K}$ mesons do not leave a measurable track in the detectors. They are measured by reconstructing their oppositely charged decay products (daughter particles). The two-body decays of \Ks create characteristic $V$-shaped particle pairs originating at the decay vertex. This topology is called $V^0$. It is searched with a dedicated $V^0$-finder algorithm that looks for track pairs of particles with opposite charges. These track pairs are extrapolated backwards until their mutual distance of the closest approach is reached. If this distance is smaller than a given limit value, the track pair becomes a $V^0$ candidate with its origin at the decay vertex. 

Two further cuts are placed on the track pairs. The first cut imposes a minimum value on the angle between the direction of the line joining the primary and decay vertices and the direction given by the vector sum of the momenta of the decay daughters. The second condition requires a minimum distance between the primary and decay vertex (a minimum length of the decaying particle). The corresponding cut requirements depend on \Ks rapidity and are listed in Methods' "Extended data" subsection. 
Starting with the approximate decay point, a $V^0$-fitter program optimizes the decay point position and the momenta of the decay daughters. Assuming that the daughter particles are pions, it is straightforward to reconstruct the invariant mass of the decaying particle (the invariant mass is defined as $m_{inv}=\sqrt{ (\sum{E_i} )^2 - ( \sum{\mathbf{p}_i} )^2}$, where $E_i$ are the energies of the decay products, $\mathbf{p}_i$ are their momenta, and $c\equiv1$ is assumed).

\begin{figure}[h!]
  \centering   
 \includegraphics[width=0.95\linewidth]{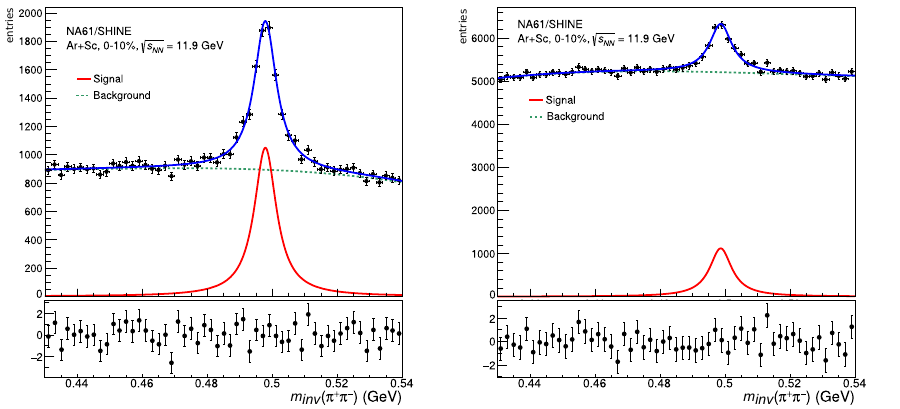}
        \vspace*{0.4cm}
        \caption{{\normalfont\bfseries Examples of fitted invariant mass distributions.}
          Two studied bins in rapidity $y$ and transverse momentum $p_T$ of the \Ks are presented,
          left: $y\in(-1.0,-0.5)$, $p_T\in(1.2,1.5)$\,\GeVc,
          right: $y\in(0.5,1.0)$, $p_T\in(1.2,1.5)$~\GeVc. Only statistical uncertainties are presented.
    The bottom panels show the difference between the experimental data and the fitted (Signal+Background) distribution, divided by the experimental uncertainty.
    }
    \label{fig:inv_mass_fit}
\end{figure}

The invariant mass distribution of $V^0$ candidates is populated by \Ks and $\Lambda$ decays, photon conversion in some detector material, and spurious particle crossings. A \Ks signal will appear as a peak on a slowly varying background. For a double-differential \Ks analysis, the momentum space was divided into seven rapidity bins ranging from $-1.5$ to 2 and nine transverse-momentum bins ranging from 0 to 2.7~\GeVc. The raw number of \Ks in a given kinematic bin is obtained from fits of appropriate signal and background functions to the invariant mass distribution of the corresponding $V^0$ candidates. The fitted signal function is taken as a Lorentzian, and the background function is a third-order Chebychev polynomial. The integral of the signal function divided by the bin width is equal to the {raw} (uncorrected) number of the reconstructed \Ks in a given kinematic bin. Two typical invariant mass distributions with signal and background fits are shown in Methods' "Extended data" subsection Fig.~\ref{fig:inv_mass_fit}.

\vspace*{0.2cm}\noindent
        {\normalfont\bfseries Corrections.}
        To correct the results for losses due to detection and data processing inefficiencies, detailed Monte Carlo simulations were performed. These simulations comprised Ar+Sc collisions generated by the EPOS model~\cite{Werner:2008zza}, and particles propagated in the \NASixtyOne detector using the GEANT framework~\cite{geant}. The charged particle tracks were reconstructed and analyzed using the same software as used for the experimental data. The branching ratio of \Ks decays was considered in the GEANT framework. The final output of the simulation consisted of reconstructed \Ks multiplicities. The ratio of the simulated and reconstructed numbers of \Ks was used as a correction factor in each $y$--\pt bin.

\begin{figure}
   \centering   
    \includegraphics[width=.8\linewidth]{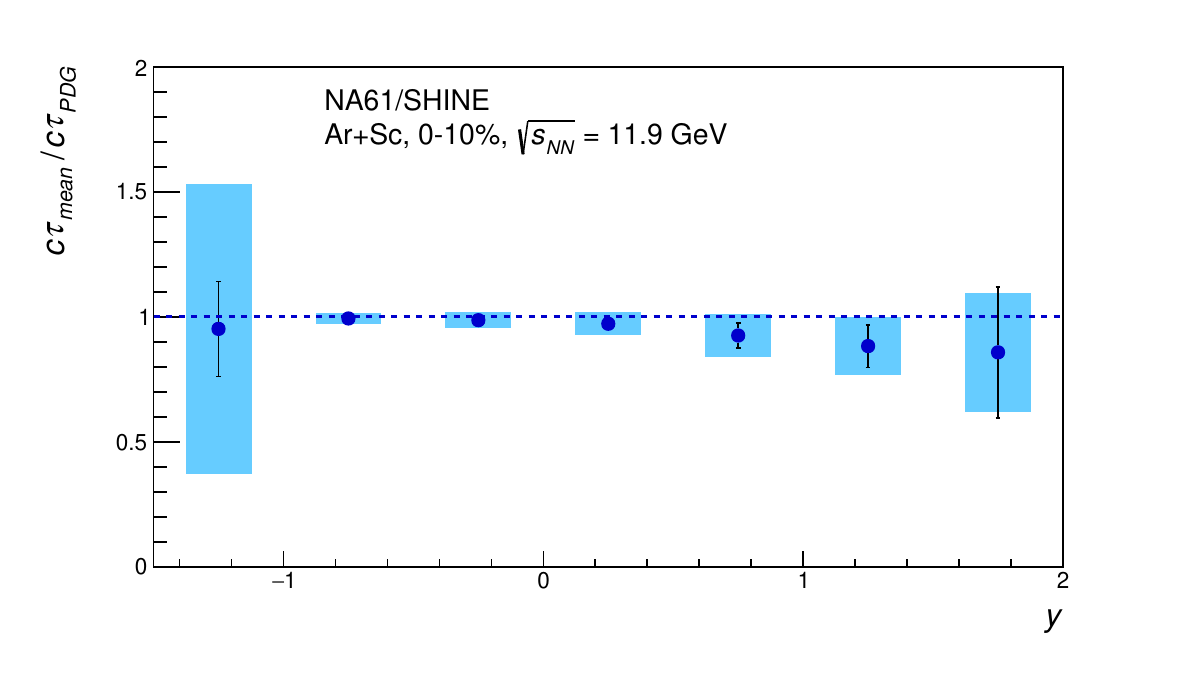}
    \caption{{\normalfont\bfseries Mean lifetime of \emph{K}$^\mathbf{0}_S$ mesons as a function of rapidity.}
    The values obtained by \NASixtyOne are divided by the PDG value~\cite{ParticleDataGroup:2024}. Statistical uncertainties are shown by vertical bars and systematic ones by shaded boxes.}
    \label{fig:lt75}
\end{figure}

        Systematic uncertainties of the measured data points were estimated by comparing the results of the entire analysis (including Monte Carlo simulations and corrections) obtained with varying cut values. The reliability of the $V^0$ reconstruction and \Ks fitting procedures can be scrutinized by studying the \Ks lifetime. Methods' "Extended data" subsection Fig.~\ref{fig:lt75} shows the computed mean lifetime of \Ks in seven rapidity bins. Good agreement with the average value provided by the PDG~\cite{ParticleDataGroup:2024} is observed.

\begin{figure}
   \centering   
     \includegraphics[width=1\linewidth]{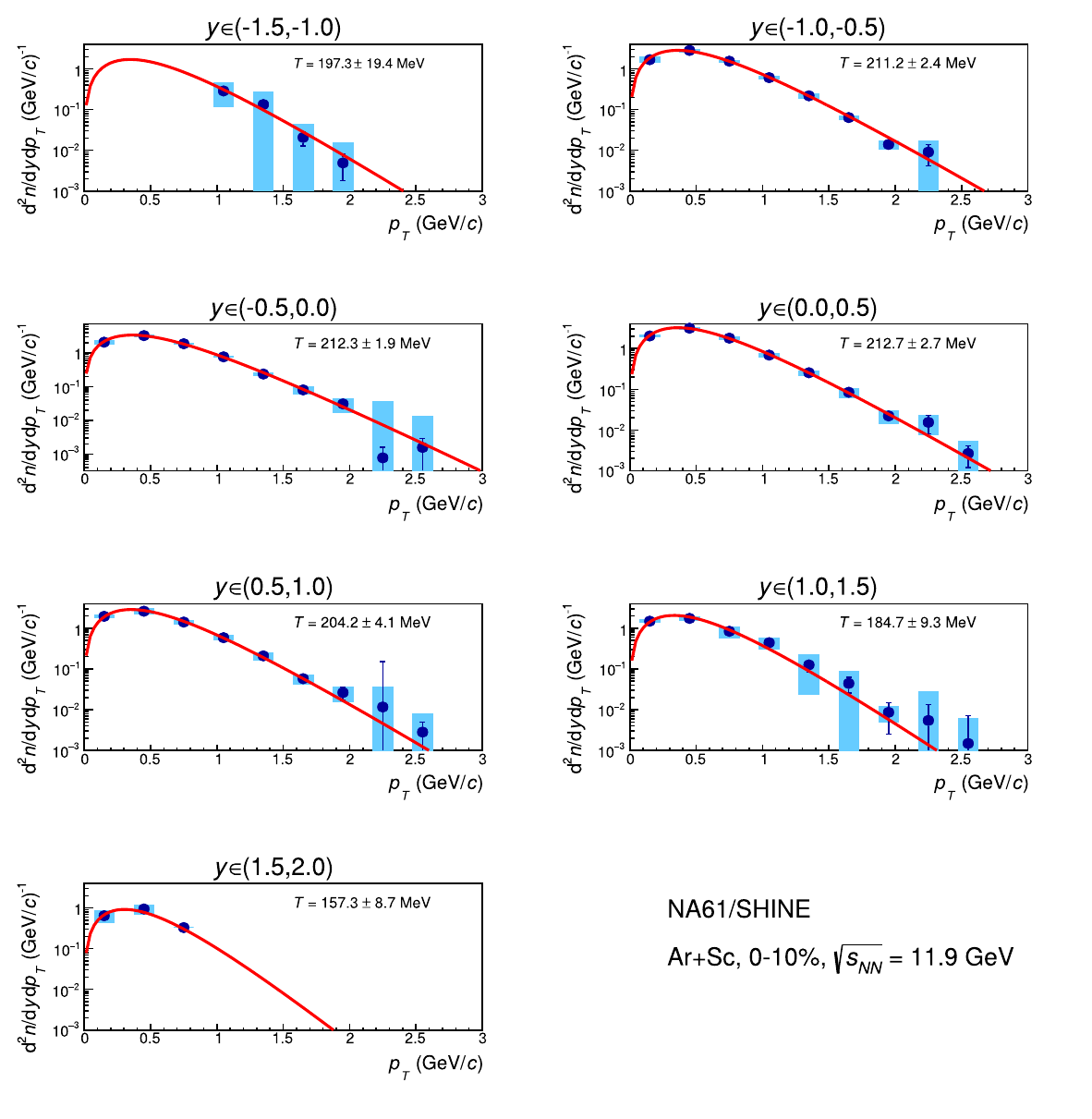}
    \caption{{\normalfont\bfseries \emph{K}$^\mathbf{0}_S$ transverse momentum spectra in rapidity bins.}
Statistical uncertainties are shown by vertical bars and systematic ones by shaded boxes. Red curves represent fits of the data with the function defined in Eq.~(\ref{eq:pt}). The inverse slope parameters ($T$), with their statistical uncertainties resulting from the fits, are also displayed inside the panels.}
    \label{fig:pt_75}
\end{figure}

\vspace*{0.2cm}\noindent
        {\normalfont\bfseries Transverse momentum distributions.}
        The distributions shown in Methods' "Extended data" subsection Fig.~\ref{fig:pt_75} represent the final results of the \Ks analysis. The \Ks yields are shown as a function of transverse momentum in seven bins of rapidity. The data points are fitted with the function:
\begin{equation}
\label{eq:pt}
    f(p_T) = A\cdot p_T \cdot \exp \left(-~\frac{\sqrt{p_T^2+m_0^2}}{T}\right),
\end{equation}
in which $A$ is a normalisation factor, $T$ is the inverse slope parameter, and $m_0$ is the \Ks  mass taken from Ref.~\cite{ParticleDataGroup:2024}. The formula assumes $c\equiv1$ for simplicity. The fit functions are plotted as red curves, and the inverse slope parameters obtained from the fits are reported in the figure legends. 
 
The transverse momentum distributions of charged and neutral $K$ mesons drawn in Fig.~\ref{fig:y_75comppt} (of the main text) are also fitted with the function defined by Eq.~(\ref{eq:pt}). 
The bottom panel of the figure presents the ratio of the two fitted curves, with its uncertainty band obtained by the propagation of the uncertainties of the fitted parameters.

\vspace*{0.2cm}\noindent
{\normalfont\bfseries Rapidity distribution.} 
The final \Ks yields in each bin of rapidity were obtained as the integrals of the curves fitted to the respective transverse momentum spectra, Eq.~(\ref{eq:pt}), including extrapolations to unmeasured regions. A comparison to the alternative method of replacing integrals in the measured regions by sums of data points only brought a negligible contribution to the systematic uncertainty. Methods' "Extended data" subsection Fig.~\ref{fig:pt_75} showed that extrapolations were needed only in the first and last rapidity bin. They amount to 88\% and 6.2\%, respectively. The large extrapolation in the first bin of rapidity increases the total uncertainty of the corresponding data point shown in Fig.~\ref{fig:y_75comp} (of the main text). In this figure, the obtained rapidity distribution of the \Ks has
been fitted with a function consisting of two Gaussians with centers displaced by a value of $\pm\Delta y$ with respect to $y=0$. These Gaussians have the same widths but may have different amplitudes. The resulting small asymmetry of the fitted rapidity distribution originates from a combined effect of the mass asymmetry of the colliding target and projectile nuclei ($A_{target} = 45$ and $A_{projectile} = 40$) and the selection of central collisions by the energy measured in the kinematic region of the projectile spectator remnants. The former favors backward and the latter forward rapidities. The yields of charged $K$ mesons at mid-rapidity listed in Table~\ref{tab1} (of the main text) were taken from Ref.~\cite{NA61SHINE:2023epu}. They were determined in the interval $0.0<y<0.2$ as discussed therein. The yield of neutral \Ks mesons at mid-rapidity was determined at $y=0$ from the aforementioned fit. Its systematic uncertainty was estimated the same way as for the data points (see above), and its statistical uncertainty was obtained by propagation of the statistical uncertainties of the fit. Both statistical and systematic uncertainties of charged and neutral $K$ yields were propagated into the ratio $R_K$. The additional uncertainty of $R_K$ resulting from the difference in the mid-rapidity definition for charged and neutral mesons was estimated to be 0.5\%, about 10\% of the total systematic uncertainty.

\vspace*{0.2cm}\noindent
{\normalfont\bfseries B. World data}

This section presents the yields of charged and neutral kaons measured by various experiments across different collision systems and energies and within specified centrality and rapidity regions. The results, presented in Table~\ref{expConditions}, are sourced directly from the original experimental publications without any modifications to ensure consistency of the quantities reported. The exceptions are HADES $K^+$ and $K^-$ yields, where two sources of systematic uncertainties were reported~\cite{HADES:2009lnd}. In our analysis, they were added in quadrature, and the square root of such a sum is shown in Table~\ref{expConditions} as the final systematic uncertainty (although in further calculations we used more precise values than 0.0014 and 0.000032 displayed in the table). In the NA49 experiment, the $K^+$ yield in Pb+Pb at $\sqrt{s_{NN}}$ $=$ 7.6 \GeV~\cite{NA49:2007stj} was reported with asymmetric systematic uncertainty; in this case, the upper limit was taken as $\sigma_{sys}$. 
For $K^+$ and $K^-$ yields in NA35 S+S collisions at $\snn$ $=$ 19.4~\GeV only statistical uncertainties were reported in the form of numerical values~\cite{NA35:1992njd}. We took the NA35 estimate of systematic uncertainty as $\thicksim$3\%~\cite{NA35:1992njd}, and the resulting numerical values are presented in the table.  
Finally, for the STAR experiment at $\snn$ $=$ 130~\GeV~\cite{STAR:2002hpr}, two types of uncertainties were reported: uncorrelated errors (first) and correlated systematic errors (second); see Ref.~\cite{STAR:2002hpr} for details.

For all kaon yields reported with statistical and systematic uncertainties separately, we calculated the total uncertainties as $\sigma_{total} = \sqrt{\sigma_{stat}^2 + \sigma_{sys}^2}$ (for STAR at $\snn$ $=$ 130~\GeV $\sigma_{total}$ was taken as $\sqrt{\sigma_{uncorr}^2 + \sigma_{corr}^2}$). Their rounded (to two significant digits) values are displayed in the third column of Table~\ref{expConditions} however, more precise values were used when propagating them to $\sigma_{total}$ of $R_K$ presented in Table~\ref{Rkvalues}. 

The notation "Yield (4$\pi$)" refers to particle mean multiplicity in full phase space. The "Yield ($y \approx 0$)" corresponds to mid-rapidity production, in most cases expressed as rapidity density $\mathrm{d}n/\mathrm{d}y$ measured in the region specified in Table~\ref{expConditions} as "\y range" (for CERES results and \NASixtyOne \Ks mesons the fits at mid-rapidity were used). In some cases, different intervals were used for charged and neutral kaons. When calculating the charged-to-neutral kaon ratio, we used the originally published results. 

The HADES data for Au+Au collisions at $\snn$ $=$ 2.4 \GeV~\cite{HADES:2017jgz, HADES:2018noy}, the FOPI data for Al+Al collisions at $\snn$ $=$ 2.7 \GeV~\cite{FOPI:2015psr, FOPI:2010ysc}, and the NA49 data for Pb+Pb collisions at $\snn$ $=$ 17.3~\GeV~\cite{NA49:2002pzu, Strabel:phd}
are excluded from this paper, as the charged and neutral kaons were measured in significantly different centrality intervals. Normalizing these results by the number of participants would introduce model dependence of the $R_{K}$ ratio. Moreover, we also omit kaon yields evaluated by the Authors of Ref.~\cite{Gazdzicki:1996pk} based on rapidity spectra measured by AGS experiments in Si+Al/Si collisions at $\snn$ $=$ 5.4~\GeV. The spectra of charged and neutral kaons were measured for different centralities~\cite{Gazdzicki:1996pk}, and the type of presented uncertainties is not clear. Finally, we also exclude NA35 kaon yields from S+Ag collisions at $\snn$ $=$ 19.4 \GeV \cite{Gazdzicki:1996pk, NA35:1994veo}. The type of uncertainties for charged kaon yields~\cite{Gazdzicki:1996pk} is not specified, and the charged and neutral kaons might have been measured for different centralities~\cite{Gazdzicki:1996pk, NA35:1994nny, NA35:1994veo}.

\vspace{1cm}

\begin{longtable}{| c | c | c | c | c | c |}
		\hline
		\multicolumn{6}{|c|}{\NASixtyOne experiment} \\
		  \hline
		  \multicolumn{6}{|c|}{Ar+Sc collisions at $\sqrt{s_{NN}}$ $=$ 11.9 \GeV} \\
		  \hline
         hadron & Yields ($y \approx 0$) $\pm$ $\sigma_{stat}$ $\pm$ $\sigma_{sys}$ & $\sigma_{total}$ & Centrality & \y ranges & Ref.\\
         \hline
		  $K^+$ & 3.732 $\pm$ 0.016 $\pm$ 0.148 & 0.15 & 0--10\% &  0.0 < $y$ < 0.2 & \cite{NA61SHINE:2023epu} \\
         \hline
         $K^-$ & 2.029 $\pm$ 0.012 $\pm$ 0.069 & 0.070 & 0--10\% &   0.0 < $y$ < 0.2 & \cite{NA61SHINE:2023epu} \\
         \hline
         $K_{S}^0$ & 2.433 $\pm$ 0.027 $\pm$ 0.102 & 0.11 & 0--10\% &  \y $=$ 0 & \footnotesize{this analysis} \\
		 \hline
      	 \multicolumn{6}{|c|}{HADES experiment} \\
		 \hline
		 \multicolumn{6}{|c|}{Ar+KCl collisions at $\sqrt{s_{NN}}$ $=$ 2.6 \GeV} \\
		 \hline
         hadron & Yields ($4\pi$) $\pm$ $\sigma_{stat}$ $\pm$ $\sigma_{sys}$ & $\sigma_{total}$  & Centrality & \y ranges & Ref. \\
      	 \hline
		 $K^{+}$ & 0.028 $\pm$ 0.002 $\pm$ 0.0014 $^{(*)}$ & 0.0024 & 0--35\% & \footnotesize{extrapolated to $4\pi$} & ~\cite{HADES:2009lnd} \\
         \hline
          $K^-$ & 0.00071 $\pm$ 0.00015 $\pm$ 0.000032 $^{(*)}$ &	0.00015 & 0--35\% & \footnotesize{extrapolated to $4\pi$} & ~\cite{HADES:2009lnd}  \\
         \hline
         $K_{S}^{0}$ & 0.0115 $\pm$ 0.0005 $\pm$ 0.0009 & 0.0010 & 0--35\% & \footnotesize{extrapolated to $4\pi$} & ~\cite{Agakishiev:2010zw}\\
		 \hline
		 \multicolumn{6}{|c|}{STAR (BES I) experiment} \\
		 \hline
		 \multicolumn{6}{|c|}{Au+Au collisions at $\sqrt{s_{NN}}$ $=$ 7.7 \GeV} \\
		 \hline
         hadron & Yields ($y \approx 0$) $\pm$ $\sigma_{stat}$ $\pm$ $\sigma_{sys}$ & $\sigma_{total}$  & Centrality & \y ranges & Ref. \\
         \hline
		 $K^+$ & 20.8 & 1.7 & 0--5\% &  $-$0.1 < $y$ < 0.1 & ~\cite{STAR:2017sal} \\
         \hline
         $K^-$ & 7.7 & 0.6 & 0--5\% &  $-$0.1 < $y$ < 0.1 & ~\cite{STAR:2017sal} \\
         \hline
         $K_{S}^0$ & 12.67$\pm$ 0.12 $\pm$ 0.44 & 0.46 & 0--5\% & $-$0.5 < $y$ < 0.5 & ~\cite{STAR:2019bjj} \\
		 \hline
		 \multicolumn{6}{|c|}{STAR (BES I) experiment} \\
		 \hline
		 \multicolumn{6}{|c|}{Au+Au collisions at $\sqrt{s_{NN}}$ $=$ 11.5 \GeV} \\
		 \hline
         hadron & Yields ($y \approx 0$) $\pm$ $\sigma_{stat}$ $\pm$ $\sigma_{sys}$ & $\sigma_{total}$  & Centrality & \y ranges & Ref. \\
         \hline
		 $K^+$ & 25.0 & 2.5 & 0--5\% & $-$0.1 < $y$ < 0.1 & ~\cite{STAR:2017sal} \\
         \hline
         $K^-$ & 12.3 & 1.2 & 0--5\% & $-$0.1 < $y$ < 0.1 & ~\cite{STAR:2017sal} \\
         \hline
         $K_{S}^0$ & 15.93 $\pm$	0.12 $\pm$	0.58 &	0.59  & 0--5\% & $-$0.5 < $y$ < 0.5 & ~\cite{STAR:2019bjj} \\
		 \hline
		 \multicolumn{6}{|c|}{STAR (BES I) experiment} \\
		 \hline
		 \multicolumn{6}{|c|}{Au+Au collisions at $\sqrt{s_{NN}}$ $=$ 19.6 \GeV} \\
		 \hline
         hadron & Yields ($y \approx 0$) $\pm$ $\sigma_{stat}$ $\pm$ $\sigma_{sys}$ & $\sigma_{total}$  & Centrality & \y ranges & Ref. \\
         \hline
		 $K^+$ & 29.6 & 2.9 & 0--5\% &  $-$0.1 < $y$ < 0.1 & ~\cite{STAR:2017sal} \\
         \hline
         $K^-$ & 18.8 &	1.9 & 0--5\% & $-$0.1 < $y$ < 0.1 & ~\cite{STAR:2017sal} \\
         \hline
         $K_{S}^0$ & 20.89 $\pm$	0.08 $\pm$ 0.67 & 0.67 & 0--5\% & $-$0.5 < $y$ < 0.5 & ~\cite{STAR:2019bjj} \\
		 \hline
		 \multicolumn{6}{|c|}{STAR (BES I) experiment} \\
		 \hline
		 \multicolumn{6}{|c|}{Au+Au collisions at $\sqrt{s_{NN}}$ $=$ 27 \GeV} \\
		 \hline
         hadron & Yields ($y \approx 0$) $\pm$ $\sigma_{stat}$ $\pm$ $\sigma_{sys}$ & $\sigma_{total}$  & Centrality & \y ranges & Ref. \\
         \hline
		 $K^+$ & 31.1 & 2.8 & 0--5\% &  $-$0.1 < $y$ < 0.1 & ~\cite{STAR:2017sal} \\
         \hline
         $K^-$ & 22.6 & 2.0 & 0--5\% & $-$0.1 < $y$ < 0.1 & ~\cite{STAR:2017sal} \\
         \hline
         $K_{S}^0$ & 23.24 $\pm$ 0.09 $\pm$ 0.70 & 0.71  & 0--5\% & $-$0.5 < $y$ < 0.5 & ~\cite{STAR:2019bjj} \\
		 \hline
		 \multicolumn{6}{|c|}{STAR (BES I) experiment} \\
		 \hline
		 \multicolumn{6}{|c|}{Au+Au collisions at $\sqrt{s_{NN}}$ $=$ 39 \GeV} \\
		 \hline
         hadron & Yields ($y \approx 0$) $\pm$ $\sigma_{stat}$ $\pm$ $\sigma_{sys}$ & $\sigma_{total}$  & Centrality & \y ranges & Ref. \\
         \hline
		 $K^+$ & 32.0 & 2.9 & 0--5\% &  $-$0.1 < $y$ < 0.1 & ~\cite{STAR:2017sal} \\
         \hline
         $K^-$ & 25.0  & 2.3 & 0--5\% & $-$0.1 < $y$ < 0.1 & ~\cite{STAR:2017sal} \\
         \hline
         $K_{S}^0$ & 24.9 $\pm$ 0.1 $\pm$ 1.7 & 1.7 & 0--5\% & $-$0.5 < $y$ < 0.5 & ~\cite{STAR:2019bjj} \\
		 \hline
      \multicolumn{6}{|c|}{NA49 experiment} \\
		 \hline
		 \multicolumn{6}{|c|}{Pb+Pb collisions at $\sqrt{s_{NN}}$ $=$ 7.6 \GeV} \\
		 \hline
         hadron & Yields ($4\pi$) $\pm$ $\sigma_{stat}$ $\pm$ $\sigma_{sys}$ & $\sigma_{total}$  & Centrality & \y ranges & Ref. \\
         \hline
		 $K^+$ & 52.9 $\pm$ 0.9 $\pm$ 3.5 $^{(*)}$ & 3.6 & 0--7.2\% & \footnotesize{extrapolated to $4\pi$} & ~\cite{NA49:2007stj} \\
         \hline
         $K^-$ & 16.0 $\pm$ 0.2 $\pm$ 0.4 & 0.45 & 0--7.2\% & \footnotesize{extrapolated to $4\pi$} & ~\cite{NA49:2007stj}  \\
         \hline
         $K_{S}^0$ & 29.3 $\pm$ 0.3 $\pm$ 2.9 & 2.9 & 0--7.2\% & \footnotesize{extrapolated to $4\pi$} & ~\cite{Strabel:phd}  \\
		 \hline
		 \multicolumn{6}{|c|}{NA49 experiment} \\
		 \hline
		 \multicolumn{6}{|c|}{Pb+Pb collisions at $\sqrt{s_{NN}}$ $=$ 8.7 \GeV} \\
		 \hline
         hadron & Yields ($4\pi$) $\pm$ $\sigma_{stat}$ $\pm$ $\sigma_{sys}$ & $\sigma_{total}$  & Centrality & \y ranges & Ref. \\
         \hline
		 $K^+$ & 59.1 $\pm$ 1.9 $\pm$ 3 & 3.6 & 0--7.2\% & \footnotesize{extrapolated to $4\pi$}  & ~\cite{NA49:2002pzu}  \\
         \hline
         $K^-$ & 19.2 $\pm$ 0.5 $\pm$ 1.0 & 1.1 & 0--7.2\% & \footnotesize{extrapolated to $4\pi$} & ~\cite{NA49:2002pzu}  \\
         \hline
         $K_{S}^0$ & 34.2 $\pm$	0.2 $\pm$ 3.4 &	3.4  & 0--7.2\% & \footnotesize{extrapolated to $4\pi$} & ~\cite{Strabel:phd}  \\
		 \hline
		 \multicolumn{6}{|c|}{CERES experiment} \\
		 \hline
		 \multicolumn{6}{|c|}{Pb+Au collisions at $\sqrt{s_{NN}}$ $=$ 17.3 \GeV} \\
		 \hline
         hadron & Yields ($y \approx 0$) $\pm$ $\sigma_{stat}$ $\pm$ $\sigma_{sys}$ & $\sigma_{total}$  & Centrality & \y ranges & Ref. \\
         \hline
		 $K^+$ &  31.8 $\pm$ 0.6 $\pm$ 2.5 & 2.6 & 0--7\% & $y$ $=$ 0 & ~\cite{Kalisky:2008ozf} \\
         \hline
         $K^-$ & 19.3 $\pm$	0.4 $\pm$ 2.0	& 2.0 & 0--7\% & $y$ $=$ 0 & ~\cite{Kalisky:2008ozf} \\
         \hline
         $K_{S}^0$ & 21.2 $\pm$ 0.9 $\pm$ 1.7 & 1.9 & 0--7\% & $y$ $=$ 0 & ~\cite{Radomski:2008zz, RadomskiPhD} \\
		 \hline
		 \multicolumn{6}{|c|}{NA35 experiment} \\
		 \hline
		 \multicolumn{6}{|c|}{S+S collisions at $\sqrt{s_{NN}}$ $=$ 19.4 \GeV} \\
		 \hline
         hadron & Yields ($4\pi$) $\pm$ $\sigma_{stat}$ $\pm$ $\sigma_{sys}$ & $\sigma_{total}$  & Centrality & \y ranges & Ref. \\
         \hline
		 $K^+$ & 12.5 $\pm$ 0.4 $\pm$ 0.375 $^{(*)}$  & 0.55 & 0--2\% & \footnotesize{extrapolated to $4\pi$}  & ~\cite{NA35:1992njd}  \\
         \hline
         $K^-$ & 6.9 $\pm$ 0.4 $\pm$ 0.207 $^{(*)}$ & 0.45 & 0--2\% & \footnotesize{extrapolated to $4\pi$} & ~\cite{NA35:1992njd}  \\
         \hline
         $K_{S}^0$ & 10.5  & 1.7 & 0--2\%  & \footnotesize{extrapolated to $4\pi$} & ~\cite{NA35:1994veo} \\
		 \hline
		 \multicolumn{6}{|c|}{STAR experiment} \\
		 \hline
		 \multicolumn{6}{|c|}{Au+Au collisions at $\sqrt{s_{NN}}$ $=$ 62.4 \GeV} \\
		 \hline
         hadron & Yields ($y \approx 0$) $\pm$ $\sigma_{stat}$ $\pm$ $\sigma_{sys}$ & $\sigma_{total}$  & Centrality & \y ranges & Ref. \\
         \hline
		  $K^+$ & 37.6 & 2.7 & 0--5\% & $-$0.1 < $y$ < 0.1 & ~\cite{STAR:2008med} \\
         \hline
         $K^-$ & 32.4 & 2.3 & 0--5\% & $-$0.1 < $y$ < 0.1 & ~\cite{STAR:2008med} \\
         \hline
         $K_{S}^0$ & 27.4 $\pm$ 0.6 $\pm$ 2.9 &	3.0  & 0--5\% & $-$1 < $y$ < 1 & ~\cite{STAR:2010yyv} \\
     \hline
      \multicolumn{6}{|c|}{STAR experiment}  \\
		 \hline
		 \multicolumn{6}{|c|}{Au+Au collisions at $\sqrt{s_{NN}}$ $=$ 130 \GeV} \\
		 \hline
         hadron & Yields ($y \approx 0$) $\pm$ $\sigma_{uncorr}$ $\pm$ $\sigma_{corr}$ & $\sigma_{total}$  & Centrality & \y ranges & Ref. \\
         \hline
		 $K^+$ &  46.2 $\pm$ 0.6 $\pm$ 6.0  & 6.0 & 0--6\% &  $^{-0.1}_{-0.5}$ <$y$ < $^{0.1}_{0.5}$ & ~\cite{STAR:2002hpr} \\
         \hline
         $K^-$ & 41.9 $\pm$ 0.6 $\pm$ 5.4  & 5.4 & 0--6\% & $^{-0.1}_{-0.5}$ < $y$ < $^{0.1}_{0.5}$ & ~\cite{STAR:2002hpr} \\
         \hline
         $K_{S}^0$ & 33.9 $\pm$ 1.1 $\pm$ 5.1  &	5.2  & 0--6\% & $-$0.5 < $y$ < 0.5 & ~\cite{STAR:2002hpr} \\
		 \hline
		 \multicolumn{6}{|c|}{STAR experiment} \\
		 \hline
		 \multicolumn{6}{|c|}{Au+Au collisions at $\sqrt{s_{NN}}$ $=$ 200 \GeV} \\
		 \hline
         hadron & Yields ($y \approx 0$) $\pm$ $\sigma_{stat}$ $\pm$ $\sigma_{sys}$ & $\sigma_{total}$  & Centrality & \y ranges & Ref. \\
         \hline
		 $K^+$ & 51.3 & 6.5 & 0--5\% & $-$0.1 < $y$ < 0.1 & ~\cite{STAR:2008med} \\
         \hline
         $K^-$ & 49.5 & 6.2 & 0--5\% & $-$0.1 < $y$ < 0.1 & ~\cite{STAR:2008med} \\
         \hline
         $K_{S}^0$ & 43.5 & 2.4 & 0--5\% & $-$0.5 < $y$ < 0.5 & ~\cite{STAR:2011fbd} \\
		 \hline
		 \multicolumn{6}{|c|}{ALICE experiment} \\
		 \hline
		 \multicolumn{6}{|c|}{Pb+Pb collisions at $\sqrt{s_{NN}}$ $=$ 2760 \GeV} \\
		 \hline
         hadron & Yields ($y \approx 0$) $\pm$ $\sigma_{stat}$ $\pm$ $\sigma_{sys}$ & $\sigma_{total}$  & Centrality & \y ranges & Ref. \\
         \hline
		 $K^+$ &  109 & 9 & 0--5\% & $-$0.5 < $y$ < 0.5 & ~\cite{ALICE:2013mez} \\
         \hline
         $K^-$ & 109 & 9 & 0--5\% & $-$0.5 < $y$ < 0.5 & ~\cite{ALICE:2013mez} \\
         \hline
         $K_{S}^0$ & 110 & 10 & 0--5\% & $-$0.5 < $y$ < 0.5 & ~\cite{ALICE:2013cdo} \\
		 \hline
    \caption{{\normalfont\bfseries 
    The compilation of world data on charged and neutral kaon yields in nucleus-nucleus collisions.} 
    The yields labeled $y \approx 0$ and $4 \pi$ correspond to the rapidity density at mid-rapidity and mean multiplicity in full phase space. The uncertainty fields are left empty in case they are not published. The systematic uncertainties labeled by~$^{(*)}$ were estimated for this analysis based on the information given in the original papers (see the text). 
    Only results corresponding to the same centrality for charged and neutral kaons are compiled.
    The "\y range" specifies the rapidity range used to obtain a given kaon yield.
    }
\label{expConditions}
\end{longtable}

Table~\ref{Rkvalues} presents the ratios of charged-to-neutral kaons from various experiments, with estimated statistical and total uncertainties where available. Taking into account the possibility of using our compilation in future analyses, the numerical values in Table~\ref{Rkvalues} are presented with unusually high precision.
\\

\begin{longtable}{| c | c | c | c | c | c | c |}
		\hline
		Experiment & Collision system & $\sqrt{s_{NN}}$ (\GeV) & $R_{K}$ & $\sigma_{stat}$ & $\sigma_{total}$ \\
		\hline
		\NASixtyOne & Ar+Sc & 11.9 & 1.1839 & 0.0138 & 0.0615 \\
		\hline
		HADES & Ar+KCl & 2.6 & 1.2483	& 0.1027 & 0.1545 \\
		\hline
		STAR (BES I) & Au+Au & 7.7 & 1.1247 & - & 0.0819 \\
		\hline
		STAR (BES I) & Au+Au & 11.5 & 1.1707 & - & 0.0973 \\
		\hline
		STAR (BES I) & Au+Au & 19.6 & 1.1584 & - & 0.0910 \\
		\hline
		STAR (BES I) & Au+Au & 27 & 1.1553	& - & 0.0819 \\
		\hline
		STAR (BES I) & Au+Au & 39 & 1.1446	& - & 0.1079 \\
		\hline
		NA49 & Pb+Pb & 7.6 & 1.1758  & 0.0198 & 0.1325  \\
		\hline
		NA49 & Pb+Pb & 8.7 & 1.1447 & 0.0295 & 0.1263 \\
		\hline
		CERES & Pb+Au & 17.3 & 1.2052	& 0.0539 & 0.1340	\\
		\hline
		NA35 & S+S & 19.4 & 0.9238 & - & 0.1533 \\
		\hline
		STAR & Au+Au & 62.4 & 1.2774 & - & 0.1525	\\
		\hline
		STAR & Au+Au & 130 & 1.2994 & - & 0.2331 \\
		\hline
		STAR & Au+Au & 200 & 1.1586 & - & 0.1214 \\
		\hline
		ALICE & Pb+Pb & 2760 & 0.9909 & - & 0.1071 \\
		\hline
     \caption{{\normalfont\bfseries Ratios of charged kaons to neutral kaons in different experiments.}
     }
    \label{Rkvalues}
\end{longtable}

{\normalfont\bfseries C. Models}

\noindent
{\normalfont\bfseries Hadron Resonance Gas model.} 
We use the Hadron Resonance Gas model implementation from Ref.~\cite{Vovchenko:2019pjl} to quantify the isospin-breaking effects and their interplay. HRG includes all hadrons and resonances with confirmed status in the PDG tables~\cite{Patrignani:2016xqp}. The PDG-listed masses, charges, lifetimes, and decay modes are used. Thus, HRG includes the isospin-symmetry violation due to masses and branching ratios of hadrons and resonances.

In HRG calculations, the exact net strangeness conservation is enforced, i.e., the calculations are done within the strangeness canonical ensemble (SCE)~\cite{BraunMunzinger:2001as,Cleymans:2016qnc}. 
The model parameters are baryo-chemical potential, $\mu_B$, temperature, $T$, volume of the system $V$, and the strangeness under-saturation parameter $\gamma_S$ (see Ref.~\cite{Rafelski:2015cxa}).
We adopt the simple parametrization of $\mu_B$ and $T$ as a function of collision energy introduced in Ref.~\cite{Poberezhnyuk:2019pxs}.
We have checked that $R_K$ is weakly sensitive to the strangeness suppression effect introduced by having the parameter $\gamma_S < 1$. Therefore, calculations are done for $\gamma_S = 1$ for simplicity.
The energy-dependent Breit-Wigner spectra~\cite{Vovchenko:2018fmh} model the resonance widths.

Figure~\ref{fig:kaon-ratio} (black line) shows the HRG predictions for $R_K$  
as a function of collision energy $\snn$ for $Q/B=0.4$. At low collision energies, $R_K < 1$ due to the enhancement of neutral kaon production caused by a larger number of neutrons than protons, \mbox{$Q/B < 1/2$}.
On the other hand, at high collision energies, HRG predicts 
$R_K\approx 1.018$ due to the mass difference between charged and neutral kaons produced directly. Finally,
$R_K\approx 1.032$ when kaon production via resonance decays is included. This is mostly due to $\phi$ decays, which strongly prefer the decay into charged kaons over the one into neutral kaons. 

We have checked that the system electric to baryon charge ratio $Q/B$ significantly affects $R_K$ up to $\snn \approx10 $~\GeV. 
At higher energies, pions dominate, and total electric and baryon charges are significantly larger than the corresponding net charges. Thus, $R_K$ becomes increasingly less sensitive to $Q/B$ with an increase in $\snn$.

We have checked that the strangeness grand-canonical ensemble and other popular parametrizations~\cite{Cleymans:2005xv,Becattini:2005xt} of the model parameters as a function of collision energy lead to quantitatively similar results for $R_K$, for $\snn\gtrsim 4$~\GeV.
The uncertainties of $R_K$ estimated by changing the parametrization of model parameters~\cite{Poberezhnyuk:2019pxs} 
are less than $1$\% for $\snn> 3$~\GeV. The effect of including light nuclei in the particle list is negligible.

A dedicated discussion for resonances $a_{0}(980)$ and $f_{0}(980)$ is
needed. For $a_{0}(980)$ three strong decay channels are reported as seen in PDG~\cite{ParticleDataGroup:2024}: $\pi \eta $, $K\overline{K}$, and $\pi \eta^{\prime}$, where the latter is phase-space suppressed because $m_{\pi} + m_{\eta^{\prime}} = 1.093 \text{ GeV} > m_{a_0} = 0.98 \pm 0.02 \text{ GeV}$. The PDG average ratio of $a_0(980)$ decay widths $\Gamma_{K\overline{K}}/\Gamma_{\pi \eta} =0.172\pm 0.019$ \cite{ParticleDataGroup:2024} implies that the $K\overline{K}$ branching ratio amounts to $\left( 15\pm 1\right)\%$:
\begin{equation}
BR(K\overline{K}) \approx \frac{\Gamma _{K\overline{K}}}{\Gamma
_{K\overline{K}}+\Gamma _{\pi \eta }+\Gamma _{\pi \eta ^{\prime }}}\approx \frac{
\Gamma _{K\overline{K}}}{\Gamma _{K\overline{K}}+\Gamma _{\pi \eta }}  
=\frac{1}{1+\frac{\Gamma _{\pi \eta }}{\Gamma _{K\overline{K}}}} \approx 0.15\pm 0.01\text{ .}
\end{equation}
For 
$f_{0}(980)$ the strong decay channels $\pi \pi $ and $K\overline{K}$ have been seen \cite{ParticleDataGroup:2024}. A list
of measurements for the $\pi \pi $ branching ratios is reported, each of
them larger than $50\%$, but no PDG average is provided. Building an
average for the $K\overline{K}$ branching ratio leads to $\left( 19\pm
2\right) \%$,  but to enhance reliability we vary the $K\overline{K}$ range between 10--40\%.
As a starting point, the HRG approach used in this work assumes
$BR(K^{+}K^{-})$ $=$ $BR(K^{0}\overline{K}^{\,0})$ for both resonances.
Taking into account the difference of the charged and neutral
kaon masses within Flatt\'{e}-like distributions~\cite%
{Flatte:1976xv,Baru:2004xg,Giacosa:2021mbz}, the ratio of charged and
neutral decay rates is about $1.1$ and remains smaller than $1.2$ when the
distribution parameters are varied within reasonable ranges (see, e.g. the
compilation in Ref.~\cite{Wu:2008hx}). We therefore recalculated the HRG
predictions assuming $20\%$ more charged than neutral kaons produced by
decays of $a_{0}^{0}(980)$ and $f_{0}(980)$. The ratio $R_{K}$ increases by
less than $0.5\%$. 
Under the extreme assumption $BR(K^{+}K^{-})$ $=$ $BR(K\overline{K})$ (no decays to neutral kaons) for both resonances, $a_{0}(980)$ and $f_{0}(980)$, and taking the upper limit for the $K\overline{K}$ branching ratio of $f_0(980)$ equal to $40\%$, $R_K$ increases by up 3\% at high collision energies.

\noindent
{\normalfont\bfseries UrQMD model.} 
The UrQMD transport model~\cite{Bass:1998ca,Bleicher:1999xi,Bleicher:2022kcu} describes $A$+$A$ collisions by explicitly propagating hadrons in phase space. 
During the propagation, rescattering among hadrons takes place. The particle production in this model happens via resonance decay or string excitation and fragmentation following the LUND model~\cite{Andersson:1983jt}. 

The gray squares in Fig.~\ref{fig:kaon-ratio} indicate the UrQMD model predictions. Here, we have considered central Au+Au collisions ($A$ $=$ 197, $Z$ $=$ 79, $Q/B \approx 0.4$). The predictions are shown within the $\snn$ range of 2.4 to 20~\GeV. At each energy, $10^4$ events are used for the analysis. 

One sees that for $\snn\lesssim 7$~\GeV, the predictions of UrQMD and HRG are similar.
At higher energies, the $R_K$ ratio in HRG is systematically higher than the one predicted by UrQMD. This is likely caused by UrQMD assuming $\phi$-meson decays to
be exactly isospin symmetric instead of taking the branching ratios from PDG.
This is the reason for showing the UrQMD predictions only up to 20~\GeV.

\vspace*{0.2cm}\noindent
{\normalfont\bfseries D. Extended data}

In this part, we present values of track pair cuts used in the analysis (Table~\ref{tab:deltaL_cut}), examples of fitted invariant mass distributions (Fig.~\ref{fig:inv_mass_fit}), mean lifetime (Fig.~\ref{fig:lt75}) and transverse momentum distributions (Fig.~\ref{fig:pt_75}) of \Ks in rapidity bins.

\begin{table}[!h]
\advance\leftskip-4cm
    \advance\rightskip-4cm
    \footnotesize
    \vglue0.2cm

    \begin{center}
    \begin{tabular}{c c c c c c c c c}
      \toprule
\multicolumn{2}{c }{rapidity bin} & $(-1.5, -1)$ & $(-1, -0.5)$ & $(-0.5, 0)$ &  $(0, 0.5)$ & $(0.5, 1)$ & $(1, 1.5)$ & $(1.5, 2)$ \\
\midrule
\multirow{2}{*}{cut value}     & cosine of angle      & $>$0.999 & $>$0.9995 & $>$0.9995 & $>$0.9995 & $>$0.9995 & $>$0.9999 & $>$0.9999 \\
& distance   & $>$5\,cm & $>$5\,cm & $>$7.5\,cm & $>$12.5\,cm & $>$12.5\,cm & $>$15\,cm & $>$12.5\,cm \\
\bottomrule
\vspace*{-0.6cm}
\end{tabular}
    \end{center}
    \caption{{\normalfont\bfseries Track pair cuts.} 
    Values of the cuts on (top row) the cosine of the angle between the line joining the primary and decay vertex and the direction of the vector sum of decay daughter momenta, and (bottom row) the distance between the primary and decay vertex.}
\label{tab:deltaL_cut}
\end{table}

\clearpage

\textbf{Data availability}\\
All data shown in plots are publicly available on the HEPdata repository (https://hepdata.net).

\vspace{0.5cm}
\textbf{Code availability}\\
The authors can provide the \NASixtyOne source code used upon reasonable request. 




\section*{Acknowledgments}

We would like to thank the CERN EP, BE, HSE and EN Departments for the
strong support of NA61/SHINE.
We also gratefully acknowledge discussions with Claudia Ahdida, Wojciech Broniowski, Marco van Leeuwen, Krzysztof Golec-Biernat,
Stanisław Mr\'owczy\'nski, Owe Philippsen, Rob Pisarski, Krishna Rajagopal, Jan Rafelski,
Jan Steinheimer, Leonardo Tinti, and Volodymyr Vovchenko for fruitful discussions and comments at various stages of the preparation of this article.

This work was supported by
the Hungarian Scientific Research Fund (grant NKFIH 138136\slash137812\slash138152 and TKP2021-NKTA-64),
the Polish Ministry of Science and Higher Education
(DIR\slash WK\slash\-2016\slash 2017\slash\-10-1, WUT ID-UB), the National Science Centre Poland (grants
2014\slash 14\slash E\slash ST2\slash 00018, 
2016\slash 21\slash D\slash ST2\slash 01983, 
2017\slash 25\slash N\slash ST2\slash 02575, 
2018\slash 29\slash N\slash ST2\slash 02595, 
2018\slash 30\slash A\slash ST2\slash 00226, 
2018\slash 31\slash G\slash ST2\slash 03910, 
2019\slash 33\slash B\slash ST2\slash 00613, 
2020\slash 39\slash O\slash ST2\slash 00277), 
the Norwegian Financial Mechanism 2014--2021 (grant 2019\slash 34\slash H\slash ST2\slash 00585),
the Polish Minister of Education and Science (contract No. 2021\slash WK\slash 10),
the Internationalization of the Jan Kochanowski University Doctoral School through the Polish Academy Agency for Academic Exchange NAWA STER No. BPI/STE/2023/1/00014,
the European Union's Horizon 2020 research and innovation programme under grant agreement No. 871072,
the Ministry of Education, Culture, Sports,
Science and Tech\-no\-lo\-gy, Japan, Grant-in-Aid for Sci\-en\-ti\-fic
Research (grants 18071005, 19034011, 19740162, 20740160 and 20039012,22H04943),
the German Research Foundation DFG (grants GA\,1480\slash8-1 and project 426579465),
the Bulgarian Ministry of Education and Science within the National
Roadmap for Research Infrastructures 2020--2027, contract No. D01-374/18.12.2020,
Serbian Ministry of Science, Technological Development and Innovation (grant
OI171002), Swiss Nationalfonds Foundation (grant 200020\-117913/1),
ETH Research Grant TH-01\,07-3, National Science Foundation grant
PHY-2013228 and the Fermi National Accelerator Laboratory (Fermilab),
a U.S. Department of Energy, Office of Science, HEP User Facility
managed by Fermi Research Alliance, LLC (FRA), acting under Contract
No. DE-AC02-07CH11359 and the IN2P3-CNRS (France),
the German Research Foundation grants GA1480/8-1, and the Alexander von Humboldt
Foundation.
\\

The data used in this article were collected before February 2022.


\vspace{0.5cm}
\textbf{Author contributions}\\
The \NASixtyOne Collaboration obtained the new experimental results presented in the paper and prepared a compilation of world data needed to calculate the charged-to-neutral kaon ratios.
The individual authors, F. Giacosa, M. Gorenstein, R. Poberezhniuk, and S. Samanta, contributed to the theoretical aspects of the paper.

\vspace{0.5cm}
\textbf{Competing interests}\\
The authors declare no competing interests.


\vspace{3cm}
{\Large The \NASixtyOne Collaboration}
\bigskip
\begin{sloppypar}

\noindent
{H.\;Adhikary~\href{https://orcid.org/0000-0002-5746-1268}{\includegraphics[height=1.7ex]{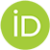}}\textsuperscript{\,13}},
{P.\;Adrich~\href{https://orcid.org/0000-0002-7019-5451}{\includegraphics[height=1.7ex]{orcid-logo.png}}\textsuperscript{\,15}},
{K.K.\;Allison~\href{https://orcid.org/0000-0002-3494-9383}{\includegraphics[height=1.7ex]{orcid-logo.png}}\textsuperscript{\,26}},
{N.\;Amin~\href{https://orcid.org/0009-0004-7572-3817}{\includegraphics[height=1.7ex]{orcid-logo.png}}\textsuperscript{\,5}},
{E.V.\;Andronov~\href{https://orcid.org/0000-0003-0437-9292}{\includegraphics[height=1.7ex]{orcid-logo.png}}\textsuperscript{\,22}}, 
{I.-C.\;Arsene~\href{https://orcid.org/0000-0003-2316-9565}{\includegraphics[height=1.7ex]{orcid-logo.png}}\textsuperscript{\,12}},
{M.\;Bajda~\href{https://orcid.org/0009-0005-8859-1099}{\includegraphics[height=1.7ex]{orcid-logo.png}}\textsuperscript{\,16}},
{Y.\;Balkova~\href{https://orcid.org/0000-0002-6957-573X}{\includegraphics[height=1.7ex]{orcid-logo.png}}\textsuperscript{\,18}},
{D.\;Battaglia~\href{https://orcid.org/0000-0002-5283-0992}{\includegraphics[height=1.7ex]{orcid-logo.png}}\textsuperscript{\,25}},
{A.\;Bazgir~\href{https://orcid.org/0000-0003-0358-0576}{\includegraphics[height=1.7ex]{orcid-logo.png}}\textsuperscript{\,13}},
{S.\;Bhosale~\href{https://orcid.org/0000-0001-5709-4747}{\includegraphics[height=1.7ex]{orcid-logo.png}}\textsuperscript{\,14}},
{M.\;Bielewicz~\href{https://orcid.org/0000-0001-8267-4874}{\includegraphics[height=1.7ex]{orcid-logo.png}}\textsuperscript{\,15}},
{A.\;Blondel~\href{https://orcid.org/0000-0002-1597-8859}{\includegraphics[height=1.7ex]{orcid-logo.png}}\textsuperscript{\,4}},
{M.\;Bogomilov~\href{https://orcid.org/0000-0001-7738-2041}{\includegraphics[height=1.7ex]{orcid-logo.png}}\textsuperscript{\,2}},
{Y.\;Bondar~\href{https://orcid.org/0000-0003-2773-9668}{\includegraphics[height=1.7ex]{orcid-logo.png}}\textsuperscript{\,13}},
{A.\;Brandin\textsuperscript{\,22}},
{W.\;Bryli\'nski~\href{https://orcid.org/0000-0002-3457-6601}{\includegraphics[height=1.7ex]{orcid-logo.png}}\textsuperscript{\,21}},
{J.\;Brzychczyk~\href{https://orcid.org/0000-0001-5320-6748}{\includegraphics[height=1.7ex]{orcid-logo.png}}\textsuperscript{\,16}},
{M.\;Buryakov~\href{https://orcid.org/0009-0008-2394-4967}{\includegraphics[height=1.7ex]{orcid-logo.png}}\textsuperscript{\,22}},
{A.F.\;Camino\textsuperscript{\,28}},
{M.\;\'Cirkovi\'c~\href{https://orcid.org/0000-0002-4420-9688}{\includegraphics[height=1.7ex]{orcid-logo.png}}\textsuperscript{\,23}},
{M.\;Csan\'ad~\href{https://orcid.org/0000-0002-3154-6925}{\includegraphics[height=1.7ex]{orcid-logo.png}}\textsuperscript{\,8}},
{J.\;Cybowska~\href{https://orcid.org/0000-0003-2568-3664}{\includegraphics[height=1.7ex]{orcid-logo.png}}\textsuperscript{\,21}},
{T.\;Czopowicz~\href{https://orcid.org/0000-0003-1908-2977}{\includegraphics[height=1.7ex]{orcid-logo.png}}\textsuperscript{\,13}},
{C.\;Dalmazzone~\href{https://orcid.org/0000-0001-6945-5845}{\includegraphics[height=1.7ex]{orcid-logo.png}}\textsuperscript{\,4}},
{N.\;Davis~\href{https://orcid.org/0000-0003-3047-6854}{\includegraphics[height=1.7ex]{orcid-logo.png}}\textsuperscript{\,14}},
{A.\;Dmitriev~\href{https://orcid.org/0000-0001-7853-0173}{\includegraphics[height=1.7ex]{orcid-logo.png}}\textsuperscript{\,22}},
{P.~von\;Doetinchem~\href{https://orcid.org/0000-0002-7801-3376}{\includegraphics[height=1.7ex]{orcid-logo.png}}\textsuperscript{\,27}},
{W.\;Dominik~\href{https://orcid.org/0000-0001-7444-9239}{\includegraphics[height=1.7ex]{orcid-logo.png}}\textsuperscript{\,19}},
{J.\;Dumarchez~\href{https://orcid.org/0000-0002-9243-4425}{\includegraphics[height=1.7ex]{orcid-logo.png}}\textsuperscript{\,4}},
{R.\;Engel~\href{https://orcid.org/0000-0003-2924-8889}{\includegraphics[height=1.7ex]{orcid-logo.png}}\textsuperscript{\,5}},
{G.A.\;Feofilov~\href{https://orcid.org/0000-0003-3700-8623}{\includegraphics[height=1.7ex]{orcid-logo.png}}\textsuperscript{\,22}},
{L.\;Fields~\href{https://orcid.org/0000-0001-8281-3686}{\includegraphics[height=1.7ex]{orcid-logo.png}}\textsuperscript{\,25}},
{Z.\;Fodor~\href{https://orcid.org/0000-0003-2519-5687}{\includegraphics[height=1.7ex]{orcid-logo.png}}\textsuperscript{\,7,20}},
{M.\;Friend~\href{https://orcid.org/0000-0003-4660-4670}{\includegraphics[height=1.7ex]{orcid-logo.png}}\textsuperscript{\,9}},
{M.\;Ga\'zdzicki~\href{https://orcid.org/0000-0002-6114-8223}{\includegraphics[height=1.7ex]{orcid-logo.png}}\textsuperscript{\,13}},
{O.\;Golosov~\href{https://orcid.org/0000-0001-6562-2925}{\includegraphics[height=1.7ex]{orcid-logo.png}}\textsuperscript{\,22}},
{V.\;Golovatyuk~\href{https://orcid.org/0009-0006-5201-0990}{\includegraphics[height=1.7ex]{orcid-logo.png}}\textsuperscript{\,22}},
{M.\;Golubeva~\href{https://orcid.org/0009-0003-4756-2449}{\includegraphics[height=1.7ex]{orcid-logo.png}}\textsuperscript{\,22}},
{K.\;Grebieszkow~\href{https://orcid.org/0000-0002-6754-9554}{\includegraphics[height=1.7ex]{orcid-logo.png}}\textsuperscript{\,21}},
{F.\;Guber~\href{https://orcid.org/0000-0001-8790-3218}{\includegraphics[height=1.7ex]{orcid-logo.png}}\textsuperscript{\,22}},
{S.N.\;Igolkin\textsuperscript{\,22}},
{S.\;Ilieva~\href{https://orcid.org/0000-0001-9204-2563}{\includegraphics[height=1.7ex]{orcid-logo.png}}\textsuperscript{\,2}},
{A.\;Ivashkin~\href{https://orcid.org/0000-0003-4595-5866}{\includegraphics[height=1.7ex]{orcid-logo.png}}\textsuperscript{\,22}},
{A.\;Izvestnyy~\href{https://orcid.org/0009-0009-1305-7309}{\includegraphics[height=1.7ex]{orcid-logo.png}}\textsuperscript{\,22}},
{N.\;Kargin\textsuperscript{\,22}},
{N.\;Karpushkin~\href{https://orcid.org/0000-0001-5513-9331}{\includegraphics[height=1.7ex]{orcid-logo.png}}\textsuperscript{\,22}},
{E.\;Kashirin~\href{https://orcid.org/0000-0001-6062-7997}{\includegraphics[height=1.7ex]{orcid-logo.png}}\textsuperscript{\,22}},
{M.\;Kie{\l}bowicz~\href{https://orcid.org/0000-0002-4403-9201}{\includegraphics[height=1.7ex]{orcid-logo.png}}\textsuperscript{\,14}},
{V.A.\;Kireyeu~\href{https://orcid.org/0000-0002-5630-9264}{\includegraphics[height=1.7ex]{orcid-logo.png}}\textsuperscript{\,22}},
{R.\;Kolesnikov~\href{https://orcid.org/0009-0006-4224-1058}{\includegraphics[height=1.7ex]{orcid-logo.png}}\textsuperscript{\,22}},
{D.\;Kolev~\href{https://orcid.org/0000-0002-9203-4739}{\includegraphics[height=1.7ex]{orcid-logo.png}}\textsuperscript{\,2}},
{Y.\;Koshio\textsuperscript{\,10}},
{V.N.\;Kovalenko~\href{https://orcid.org/0000-0001-6012-6615}{\includegraphics[height=1.7ex]{orcid-logo.png}}\textsuperscript{\,22}},
{S.\;Kowalski~\href{https://orcid.org/0000-0001-9888-4008}{\includegraphics[height=1.7ex]{orcid-logo.png}}\textsuperscript{\,18}},
{B.\;Koz{\l}owski~\href{https://orcid.org/0000-0001-8442-2320}{\includegraphics[height=1.7ex]{orcid-logo.png}}\textsuperscript{\,21}},
{A.\;Krasnoperov~\href{https://orcid.org/0000-0002-1425-2861}{\includegraphics[height=1.7ex]{orcid-logo.png}}\textsuperscript{\,22}},
{W.\;Kucewicz~\href{https://orcid.org/0000-0002-2073-711X}{\includegraphics[height=1.7ex]{orcid-logo.png}}\textsuperscript{\,17}},
{M.\;Kuchowicz~\href{https://orcid.org/0000-0003-3174-585X}{\includegraphics[height=1.7ex]{orcid-logo.png}}\textsuperscript{\,20}},
{M.\;Kuich~\href{https://orcid.org/0000-0002-6507-8699}{\includegraphics[height=1.7ex]{orcid-logo.png}}\textsuperscript{\,19}},
{A.\;Kurepin~\href{https://orcid.org/0000-0002-1851-4136}{\includegraphics[height=1.7ex]{orcid-logo.png}}\textsuperscript{\,22}},
{A.\;L\'aszl\'o~\href{https://orcid.org/0000-0003-2712-6968}{\includegraphics[height=1.7ex]{orcid-logo.png}}\textsuperscript{\,7}},
{M.\;Lewicki~\href{https://orcid.org/0000-0002-8972-3066}{\includegraphics[height=1.7ex]{orcid-logo.png}}\textsuperscript{\,20}},
{G.\;Lykasov~\href{https://orcid.org/0000-0002-1544-6959}{\includegraphics[height=1.7ex]{orcid-logo.png}}\textsuperscript{\,22}},
{V.V.\;Lyubushkin~\href{https://orcid.org/0000-0003-0136-233X}{\includegraphics[height=1.7ex]{orcid-logo.png}}\textsuperscript{\,22}},
{M.\;Ma\'ckowiak-Paw{\l}owska~\href{https://orcid.org/0000-0003-3954-6329}{\includegraphics[height=1.7ex]{orcid-logo.png}}\textsuperscript{\,21}},
{Z.\;Majka~\href{https://orcid.org/0000-0003-3064-6577}{\includegraphics[height=1.7ex]{orcid-logo.png}}\textsuperscript{\,16}}\textsuperscript{\dag},
{A.\;Makhnev~\href{https://orcid.org/0009-0002-9745-1897}{\includegraphics[height=1.7ex]{orcid-logo.png}}\textsuperscript{\,22}},
{B.\;Maksiak~\href{https://orcid.org/0000-0002-7950-2307}{\includegraphics[height=1.7ex]{orcid-logo.png}}\textsuperscript{\,15}},
{A.I.\;Malakhov~\href{https://orcid.org/0000-0001-8569-8409}{\includegraphics[height=1.7ex]{orcid-logo.png}}\textsuperscript{\,22}},
{A.\;Marcinek~\href{https://orcid.org/0000-0001-9922-743X}{\includegraphics[height=1.7ex]{orcid-logo.png}}\textsuperscript{\,14}},
{A.D.\;Marino~\href{https://orcid.org/0000-0002-1709-538X}{\includegraphics[height=1.7ex]{orcid-logo.png}}\textsuperscript{\,26}},
{H.-J.\;Mathes~\href{https://orcid.org/0000-0002-0680-040X}{\includegraphics[height=1.7ex]{orcid-logo.png}}\textsuperscript{\,5}},
{T.\;Matulewicz~\href{https://orcid.org/0000-0003-2098-1216}{\includegraphics[height=1.7ex]{orcid-logo.png}}\textsuperscript{\,19}},
{V.\;Matveev~\href{https://orcid.org/0000-0002-2745-5908}{\includegraphics[height=1.7ex]{orcid-logo.png}}\textsuperscript{\,22}},
{G.L.\;Melkumov~\href{https://orcid.org/0009-0004-2074-6755}{\includegraphics[height=1.7ex]{orcid-logo.png}}\textsuperscript{\,22}},
{A.\;Merzlaya~\href{https://orcid.org/0000-0002-6553-2783}{\includegraphics[height=1.7ex]{orcid-logo.png}}\textsuperscript{\,12}},
{{\L}.\;Mik~\href{https://orcid.org/0000-0003-2712-6861}{\includegraphics[height=1.7ex]{orcid-logo.png}}\textsuperscript{\,17}},
{S.\;Morozov~\href{https://orcid.org/0000-0002-6748-7277}{\includegraphics[height=1.7ex]{orcid-logo.png}}\textsuperscript{\,22}},
{Y.\;Nagai~\href{https://orcid.org/0000-0002-1792-5005}{\includegraphics[height=1.7ex]{orcid-logo.png}}\textsuperscript{\,8}},
{T.\;Nakadaira~\href{https://orcid.org/0000-0003-4327-7598}{\includegraphics[height=1.7ex]{orcid-logo.png}}\textsuperscript{\,9}},
{M.\;Naskret~\href{https://orcid.org/0000-0002-5634-6639}{\includegraphics[height=1.7ex]{orcid-logo.png}}\textsuperscript{\,20}},
{S.\;Nishimori~\href{https://orcid.org/~0000-0002-1820-0938}{\includegraphics[height=1.7ex]{orcid-logo.png}}\textsuperscript{\,9}},
{A.\;Olivier~\href{https://orcid.org/0000-0003-4261-8303}{\includegraphics[height=1.7ex]{orcid-logo.png}}\textsuperscript{\,25}},
{V.\;Ozvenchuk~\href{https://orcid.org/0000-0002-7821-7109}{\includegraphics[height=1.7ex]{orcid-logo.png}}\textsuperscript{\,14}},
{O.\;Panova~\href{https://orcid.org/0000-0001-5039-7788}{\includegraphics[height=1.7ex]{orcid-logo.png}}\textsuperscript{\,13}},
{V.\;Paolone~\href{https://orcid.org/0000-0003-2162-0957}{\includegraphics[height=1.7ex]{orcid-logo.png}}\textsuperscript{\,28}},
{O.\;Petukhov~\href{https://orcid.org/0000-0002-8872-8324}{\includegraphics[height=1.7ex]{orcid-logo.png}}\textsuperscript{\,22}},
{I.\;Pidhurskyi~\href{https://orcid.org/0000-0001-9916-9436}{\includegraphics[height=1.7ex]{orcid-logo.png}}\textsuperscript{\,13}},
{R.\;P{\l}aneta~\href{https://orcid.org/0000-0001-8007-8577}{\includegraphics[height=1.7ex]{orcid-logo.png}}\textsuperscript{\,16}},
{P.\;Podlaski~\href{https://orcid.org/0000-0002-0232-9841}{\includegraphics[height=1.7ex]{orcid-logo.png}}\textsuperscript{\,19}},
{B.A.\;Popov~\href{https://orcid.org/0000-0001-5416-9301}{\includegraphics[height=1.7ex]{orcid-logo.png}}\textsuperscript{\,22,4}},
{B.\;P\'orfy~\href{https://orcid.org/0000-0001-5724-9737}{\includegraphics[height=1.7ex]{orcid-logo.png}}\textsuperscript{\,7,8}},
{D.S.\;Prokhorova~\href{https://orcid.org/0000-0003-3726-9196}{\includegraphics[height=1.7ex]{orcid-logo.png}}\textsuperscript{\,22}},
{D.\;Pszczel~\href{https://orcid.org/0000-0002-4697-6688}{\includegraphics[height=1.7ex]{orcid-logo.png}}\textsuperscript{\,15}},
{S.\;Pu{\l}awski~\href{https://orcid.org/0000-0003-1982-2787}{\includegraphics[height=1.7ex]{orcid-logo.png}}\textsuperscript{\,18}},
{J.\;Puzovi\'c\textsuperscript{\,23}\textsuperscript{\dag}},
{R.\;Renfordt~\href{https://orcid.org/0000-0002-5633-104X}{\includegraphics[height=1.7ex]{orcid-logo.png}}\textsuperscript{\,18}},
{L.\;Ren~\href{https://orcid.org/0000-0003-1709-7673}{\includegraphics[height=1.7ex]{orcid-logo.png}}\textsuperscript{\,26}},
{V.Z.\;Reyna~Ortiz~\href{https://orcid.org/0000-0002-7026-8198}{\includegraphics[height=1.7ex]{orcid-logo.png}}\textsuperscript{\,13}},
{D.\;R\"ohrich\textsuperscript{\,11}},
{E.\;Rondio~\href{https://orcid.org/0000-0002-2607-4820}{\includegraphics[height=1.7ex]{orcid-logo.png}}\textsuperscript{\,15}},
{M.\;Roth~\href{https://orcid.org/0000-0003-1281-4477}{\includegraphics[height=1.7ex]{orcid-logo.png}}\textsuperscript{\,5}},
{{\L}.\;Rozp{\l}ochowski~\href{https://orcid.org/0000-0003-3680-6738}{\includegraphics[height=1.7ex]{orcid-logo.png}}\textsuperscript{\,14}},
{B.T.\;Rumberger~\href{https://orcid.org/0000-0002-4867-945X}{\includegraphics[height=1.7ex]{orcid-logo.png}}\textsuperscript{\,26}},
{M.\;Rumyantsev~\href{https://orcid.org/0000-0001-8233-2030}{\includegraphics[height=1.7ex]{orcid-logo.png}}\textsuperscript{\,22}},
{A.\;Rustamov~\href{https://orcid.org/0000-0001-8678-6400}{\includegraphics[height=1.7ex]{orcid-logo.png}}\textsuperscript{\,1}},
{M.\;Rybczynski~\href{https://orcid.org/0000-0002-3638-3766}{\includegraphics[height=1.7ex]{orcid-logo.png}}\textsuperscript{\,13}},
{A.\;Rybicki~\href{https://orcid.org/0000-0003-3076-0505}{\includegraphics[height=1.7ex]{orcid-logo.png}}\textsuperscript{\,14}},
{D.\;Rybka\textsuperscript{\,15}},
{K.\;Sakashita~\href{https://orcid.org/0000-0003-2602-7837}{\includegraphics[height=1.7ex]{orcid-logo.png}}\textsuperscript{\,9}},
{K.\;Schmidt~\href{https://orcid.org/0000-0002-0903-5790}{\includegraphics[height=1.7ex]{orcid-logo.png}}\textsuperscript{\,18}},
{A.Yu.\;Seryakov~\href{https://orcid.org/0000-0002-5759-5485}{\includegraphics[height=1.7ex]{orcid-logo.png}}\textsuperscript{\,22}},
{P.\;Seyboth~\href{https://orcid.org/0000-0002-4821-6105}{\includegraphics[height=1.7ex]{orcid-logo.png}}\textsuperscript{\,13}},
{U.A.\;Shah~\href{https://orcid.org/0000-0002-9315-1304}{\includegraphics[height=1.7ex]{orcid-logo.png}}\textsuperscript{\,13}},
{Y.\;Shiraishi\textsuperscript{\,10}},
{A.\;Shukla~\href{https://orcid.org/0000-0003-3839-7229}{\includegraphics[height=1.7ex]{orcid-logo.png}}\textsuperscript{\,27}},
{M.\;S{\l}odkowski~\href{https://orcid.org/0000-0003-0463-2753}{\includegraphics[height=1.7ex]{orcid-logo.png}}\textsuperscript{\,21}},
{P.\;Staszel~\href{https://orcid.org/0000-0003-4002-1626}{\includegraphics[height=1.7ex]{orcid-logo.png}}\textsuperscript{\,16}},
{G.\;Stefanek~\href{https://orcid.org/0000-0001-6656-9177}{\includegraphics[height=1.7ex]{orcid-logo.png}}\textsuperscript{\,13}},
{J.\;Stepaniak~\href{https://orcid.org/0000-0003-2064-9870}{\includegraphics[height=1.7ex]{orcid-logo.png}}\textsuperscript{\,15}},
{M.\;Strikhanov\textsuperscript{\,22}},
{H.\;Str\"obele\textsuperscript{\,6}},
{T.\;\v{S}u\v{s}a~\href{https://orcid.org/0000-0001-7430-2552}{\includegraphics[height=1.7ex]{orcid-logo.png}}\textsuperscript{\,3}},
{{\L}.\;\'Swiderski~\href{https://orcid.org/0000-0001-5857-2085}{\includegraphics[height=1.7ex]{orcid-logo.png}}\textsuperscript{\,15}},
{J.\;Szewi\'nski~\href{https://orcid.org/0000-0003-2981-9303}{\includegraphics[height=1.7ex]{orcid-logo.png}}\textsuperscript{\,15}},
{R.\;Szukiewicz~\href{https://orcid.org/0000-0002-1291-4040}{\includegraphics[height=1.7ex]{orcid-logo.png}}\textsuperscript{\,20}},
{A.\;Taranenko~\href{https://orcid.org/0000-0003-1737-4474}{\includegraphics[height=1.7ex]{orcid-logo.png}}\textsuperscript{\,22}},
{A.\;Tefelska~\href{https://orcid.org/0000-0002-6069-4273}{\includegraphics[height=1.7ex]{orcid-logo.png}}\textsuperscript{\,21}},
{D.\;Tefelski~\href{https://orcid.org/0000-0003-0802-2290}{\includegraphics[height=1.7ex]{orcid-logo.png}}\textsuperscript{\,21}},
{V.\;Tereshchenko\textsuperscript{\,22}},
{R.\;Tsenov~\href{https://orcid.org/0000-0002-1330-8640}{\includegraphics[height=1.7ex]{orcid-logo.png}}\textsuperscript{\,2}},
{L.\;Turko~\href{https://orcid.org/0000-0002-5474-8650}{\includegraphics[height=1.7ex]{orcid-logo.png}}\textsuperscript{\,20}},
{T.S.\;Tveter~\href{https://orcid.org/0009-0003-7140-8644}{\includegraphics[height=1.7ex]{orcid-logo.png}}\textsuperscript{\,12}},
{M.\;Unger~\href{https://orcid.org/0000-0002-7651-0272~}{\includegraphics[height=1.7ex]{orcid-logo.png}}\textsuperscript{\,5}},
{M.\;Urbaniak~\href{https://orcid.org/0000-0002-9768-030X}{\includegraphics[height=1.7ex]{orcid-logo.png}}\textsuperscript{\,18}},
{F.F.\;Valiev~\href{https://orcid.org/0000-0001-5130-5603}{\includegraphics[height=1.7ex]{orcid-logo.png}}\textsuperscript{\,22}},
{D.\;Veberi\v{c}~\href{https://orcid.org/0000-0003-2683-1526}{\includegraphics[height=1.7ex]{orcid-logo.png}}\textsuperscript{\,5}},
{V.V.\;Vechernin~\href{https://orcid.org/0000-0003-1458-8055}{\includegraphics[height=1.7ex]{orcid-logo.png}}\textsuperscript{\,22}},
{O.\;Vitiuk~\href{https://orcid.org/0000-0002-9744-3937}{\includegraphics[height=1.7ex]{orcid-logo.png}}\textsuperscript{\,20}},
{V.\;Volkov~\href{https://orcid.org/0000-0002-4785-7517}{\includegraphics[height=1.7ex]{orcid-logo.png}}\textsuperscript{\,22}},
{A.\;Wickremasinghe~\href{https://orcid.org/0000-0002-5325-0455}{\includegraphics[height=1.7ex]{orcid-logo.png}}\textsuperscript{\,24}},
{K.\;Witek~\href{https://orcid.org/0009-0004-6699-1895}{\includegraphics[height=1.7ex]{orcid-logo.png}}\textsuperscript{\,17}},
{K.\;W\'ojcik~\href{https://orcid.org/0000-0002-8315-9281}{\includegraphics[height=1.7ex]{orcid-logo.png}}\textsuperscript{\,18}},
{O.\;Wyszy\'nski~\href{https://orcid.org/0000-0002-6652-0450}{\includegraphics[height=1.7ex]{orcid-logo.png}}\textsuperscript{\,13}},
{A.\;Zaitsev~\href{https://orcid.org/0000-0003-4711-9925}{\includegraphics[height=1.7ex]{orcid-logo.png}}\textsuperscript{\,22}},
{E.\;Zherebtsova~\href{https://orcid.org/0000-0002-1364-0969}{\includegraphics[height=1.7ex]{orcid-logo.png}}\textsuperscript{\,20}},
{E.D.\;Zimmerman~\href{https://orcid.org/0000-0002-6394-6659}{\includegraphics[height=1.7ex]{orcid-logo.png}}\textsuperscript{\,26}},
{A.\;Zviagina~\href{https://orcid.org/0009-0007-5211-6493}{\includegraphics[height=1.7ex]{orcid-logo.png}}\textsuperscript{\,22}}, and
{R.\;Zwaska~\href{https://orcid.org/0000-0002-4889-5988}{\includegraphics[height=1.7ex]{orcid-logo.png}}\textsuperscript{\,24}}

\vspace{1cm}
{\Large and individual authors}
\bigskip

\noindent
{F.\;Giacosa~\href{https://orcid.org/0000-0002-7290-9366}{\includegraphics[height=1.7ex]{orcid-logo.png}}\textsuperscript{\,13,6}},
{M.\;Gorenstein~\href{https://orcid.org/0000-0003-3032-7859}{\includegraphics[height=1.7ex]{orcid-logo.png}}\textsuperscript{\,29,30}},
{R.\;Poberezhniuk~\href{https://orcid.org/0000-0002-5559-3718}{\includegraphics[height=1.7ex]{orcid-logo.png}}\textsuperscript{\,29,30,31}},
{S.\;Samanta~\href{https://orcid.org/0000-0003-2065-9219}{\includegraphics[height=1.7ex]{orcid-logo.png}}\textsuperscript{\,32}}\vspace{1cm}
\\\rule{2cm}{.5pt}\\[-.5ex]\textit{\textsuperscript{\dag} \footnotesize deceased}

\end{sloppypar}


\noindent
\textsuperscript{1}~National Nuclear Research Center, Baku, Azerbaijan\\
\textsuperscript{2}~Faculty of Physics, University of Sofia, Sofia, Bulgaria\\
\textsuperscript{3}~Ruder Bo\v{s}kovi\'c Institute, Zagreb, Croatia\\
\textsuperscript{4}~LPNHE, Sorbonne University, CNRS/IN2P3, Paris, France\\
\textsuperscript{5}~Karlsruhe Institute of Technology, Karlsruhe, Germany\\
\textsuperscript{6}~University of Frankfurt, Frankfurt, Germany\\
\textsuperscript{7}~HUN-REN Wigner Research Centre for Physics, Budapest, Hungary\\
\textsuperscript{8}~E\"otv\"os Lor\'and University, Budapest, Hungary\\
\textsuperscript{9}~Institute for Particle and Nuclear Studies, Tsukuba, Japan\\
\textsuperscript{10}~Okayama University, Japan\\
\textsuperscript{11}~University of Bergen, Bergen, Norway\\
\textsuperscript{12}~University of Oslo, Oslo, Norway\\
\textsuperscript{13}~Jan Kochanowski University, Kielce, Poland\\
\textsuperscript{14}~Institute of Nuclear Physics, Polish Academy of Sciences, Cracow, Poland\\
\textsuperscript{15}~National Centre for Nuclear Research, Warsaw, Poland\\
\textsuperscript{16}~Jagiellonian University, Cracow, Poland\\
\textsuperscript{17}~AGH University of Krakow, Cracow, Poland\\
\textsuperscript{18}~University of Silesia, Katowice, Poland\\
\textsuperscript{19}~University of Warsaw, Warsaw, Poland\\
\textsuperscript{20}~University of Wroc{\l}aw,  Wroc{\l}aw, Poland\\
\textsuperscript{21}~Warsaw University of Technology, Warsaw, Poland\\
\textsuperscript{22}~Affiliated with an institution covered by a cooperation agreement with CERN\\
\textsuperscript{23}~University of Belgrade, Belgrade, Serbia\\
\textsuperscript{24}~Fermilab, Batavia, USA\\
\textsuperscript{25}~University of Notre Dame, Notre Dame, USA\\
\textsuperscript{26}~University of Colorado, Boulder, USA\\
\textsuperscript{27}~University of Hawaii at Manoa, Honolulu, USA\\
\textsuperscript{28}~University of Pittsburgh, Pittsburgh, USA\\
\textsuperscript{29}~Bogolyubov Institute for Theoretical Physics, Kyiv, Ukraine\\
\textsuperscript{30}~Frankfurt Institute for Advanced Studies, Giersch Science Center, Frankfurt am Main, Germany\\
\textsuperscript{31}~University of Houston, Houston, USA\\
\textsuperscript{32}~School of Applied Sciences, Kalinga Institute of Industrial Technology, Bhubaneswar, Odisha, India\\


\newpage

\end{document}